\documentclass[fleqn,usenatbib]{mnras}

\usepackage{newtxtext} 

\usepackage[T1]{fontenc}
\usepackage{ae,aecompl}

\usepackage{graphicx}	
\usepackage{amsmath}	
\usepackage{amssymb}	

\newcommand{\Rsun}{$\mathrm{R}_\odot$}
\newcommand{\Ni}{$^{56}$Ni}

\newcommand{\appropto}{\mathrel{\vcenter{
  \offinterlineskip\halign{\hfil$##$\cr
    \propto\cr\noalign{\kern2pt}\sim\cr\noalign{\kern-2pt}}}}}

\title[Treatment of line opacity and SN light curves]
{The influence of line opacity treatment in STELLA on supernova light curves}

\author[A. Kozyreva et al.]{Alexandra~Kozyreva$^{\,1,2}$\thanks{E-mail: sasha@mpa-garching.mpg.de},
Luke Shingles$^{\,3}$,
Alexey Mironov$^{\,4}$
\newauthor
Petr Baklanov$^{\,5,6,7}$
Sergey Blinnikov$^{\,4,5,6,8}$
\\
$^{1}$Max-Planck-Institut f\"ur Astrophysik, Garching bei M\"unchen, 85748, Germany,\\
$^{2}$Alexander von Humboldt Fellowship\\
$^{3}$Astrophysics Research Centre, School of Mathematics and Physics, Queen's University, Belfast BT7 1NN, UK\\
$^{4}$Sternberg astronomical institute of Lomonosov Moscow State University, 119992, Moscow, Russia\\
$^{5}$Space Research Institute (IKI), 84/32 Profsoyuznaya, Moscow, 117997, Russia\\
$^{6}$NRC ``Kurchatov Institute'' -- ITEP, Moscow, 117218, Russia \\
$^{7}$National Research Nuclear University Moscow Engineering Physics Institute, Moscow, 115409, Russia\\
$^{8}$ Dukhov Automatics Research Institute (VNIIA), 127055 Moscow, 127055, Russia\\
}

\date{Accepted XXX. Received YYY; in original form ZZZ}

\pubyear{2020}

\begin{document}
\label{firstpage}
\pagerange{\pageref{firstpage}--\pageref{lastpage}}
\maketitle

\begin{abstract}
We systematically explore the effect of the treatment of line opacity on
supernova light curves. We find that it is important to consider line
opacity for both scattering and absorption (i.e. thermalisation which mimics the effect of fluorescence.)
We explore the impact of degree of thermalisation
on three major types of supernovae: Type\,Ia, Type~II-peculiar,
and Type\,II-plateau. For that we use radiative transfer code STELLA and analyse
broad-band light curves in the context of simulations done with the spectral
synthesis code ARTIS and in the context a few examples of observed supernovae
of each type. We found that the plausible range for the
ratio between absorption and scattering in the radiation hydrodynamics code
STELLA is (0.8-1):(0.2--0), i.e. the recommended thermalisation parameter
is 0.9.

\end{abstract}




\section[Motivation]{Motivation}
\label{sect:intro}

Large sets of observational data have become available from
 supernova-search surveys and transient robotic systems.
Among these data, there are dozens of discovered supernovae (SNe) with spectral snapshots spanning the
earliest epochs up to hundreds of days after explosion. The comparison between observed spectral evolution of SNe and numerical simulations
provides clues about the progenitor systems and the explosion mechanisms.
One of the tasks for theoretical studies using radiative
transfer simulations is to reproduce the evolution of the radiation field in the
fast moving SN ejecta and the energy distribution across a wide spectral
range. A number of sophisticated radiative transfer codes in the literature
are used to carry out detailed spectral synthesis simulations
\citep[][ and many
others]{1993A&A...279..447M,1996MNRAS.278..763B,2006ApJ...651..366K,2010MNRAS.405.2141D,2011A&A...530A..45J,2013ApJS..209...36W}.
Other codes do not simulate spectra but assume approxiate treatments for
the formation of lines and the redistribution of energy resulting from
radiation--matter interaction \citep[][ and
others]{1998ApJ...496..454B,2011ApJ...729...61B,2014ApJ...792L..11P,2015AandA...581A..40U}.

In the current study, we discuss a particular aspect of radiative transfer simulations for
SNe that reflects the microphysics of photon--atom interaction, namely the
probability for photons to be either resonantly scattered or
inelastically absorbed.
For the sophisticated codes from the first group, there is no
simplification assumed.
In the second group of codes, these processes are treated approximately with a thermalisation parameter
that determines the ratio of scattering to absorption.
In this paper, we discuss the
best value to use for the thermalisation parameter in the hydrodynamics radiative
transfer code \verb|STELLA|{} by comparison to observed SNe and to advanced spectral synthesis codes.

\begin{table*}
\caption{The choice of thermalisation parameter in different radiative transfer
codes. Note, that \citet{1996MNRAS.283..297B} apply $\varepsilon$ to a
series of lines considered in LTE while doing full non-LTE
radiative transfer. \citet{2018ApJ...852L..33G} apply $\varepsilon=0$ for
elements with atomic number $Z\le20$, and $\varepsilon=1$ for $Z>20$. See
text below for details about the choice of the thermalisation parameter in
\citet{1998ApJ...496..454B}.}
\label{table:values}
\begin{tabular}{l|c|c|c|l|l}
Reference                  & Absorption & Scattering & Code name & Application & Details\\
\hline
\citet{1996MNRAS.283..297B}   & 0.05--0.1 & 0.9-0.95 & PHOENIX   & SNe~Ia, II & Non-LTE with a number of LTE lines \\
\citet{1997ApJ...485..812N}   & 0.1       & 0.9      & PHOENIX   & SNe~Ia     & as above\\
\citet{1998ApJ...496..454B}   & 0/1       & 1/0      & STELLA    & SNe~I, II  & LTE; no temperature for radiation \\
\citet{2009MNRAS.398.1809K}   & --        & --       & ARTIS     & SNe~I, II  &   \\
\citet{2010MNRAS.405.2141D}   & --        & --       & CMFGEN    & SNe~I, II  & full Non-LTE from kinetic equations \\
\citet{2006ApJ...651..366K}   & 0.3-1     & 0.7-0    & SEDONA    & SNe~Ia, II & LTE \\
\citet{2018ApJ...852L..33G}   & 1/0       & 0/1      & SEDONA    & SNe~Ia     & as above \\
\citet{2018ApJ...854...52S}   & 1         & 0        & SEDONA    & SNe~Ia     & as above \\
\citet{2015AandA...581A..40U} & 0         & 1        & CRAB      & SNe~II     & LTE with corrections for Non-LTE, grey atmosphere \\
\citet{2018AandA...614A.115M} & 0.9       & 0.1      & TURTLS    & early SNe~Ia & LTE \\
\end{tabular}
\end{table*}

In Table~\ref{table:values}, we list several radiative transfer codes
and their assumed values for the thermalisation parameter. Note that a few papers
using \verb|SEDONA|{} state different values of the thermalisation parameter, while
the standard default value is $\varepsilon=0.9$ for recent studies (Nathaniel Roth, Daniel Kasen, private
communication).
The choice of pure absorption line opacity can be justified on the basis that the strongest lines
are iron lines, and detailed studies \citep[e.g.][]{2006ApJ...649..939K} show that the iron
lines are mostly purely absorptive. Although photons are not immediately
thermalised, the true thermalisation timescale is extremely short
\citep{2000ApJ...530..757P}. Initial high-energy photons are absorbed and
re-emitted at longer wavelengths, i.e. the effect of fluorescence occurs broadly in the SN ejecta
\cite[e.g.][]{1995ApJ...443...89H,1999A&A...345..211L,2000ApJ...530..757P}.
\citet{2018ApJ...852L..33G} suggest treating all
lines from elements with atomic numbers below 20 ($Z\le20$) as ``purely scattering'' and all
lines from elements with atomic numbers above 20 as ``purely absorptive''.
\citet{2006ApJ...649..939K} analysed the effect of Ca\,II triplet
on \emph{I} band light curve and concluded that Ca must be treated as purely
scattering, otherwise the \emph{I} magnitude does not match observed light
curves, in particular at the second maximum.

In the present study, we address the accuracy of line opacity treatment in
the hydrodynamics radiative transfer code \verb|STELLA|.
In the basic descriptive paper about \verb|STELLA|{}, the authors raised
the issue of the choice between scattering and absorptive
treatment of lines \citep{1998ApJ...496..454B}. The \verb|STELLA|{} light
curves were compared to those calculated with \verb|EDDINGTON|{}
\citep{1993ApJ...412..731E}, which is a full Non-LTE (no assumed Local Thermodynamic Equilibrium) radiative transfer code.
Note that simulations with the LTE option and forced absorptive line opacity
in \verb|EDDINGTON|{} were used for that comparison analysis.
\citet{1998ApJ...496..454B} concluded that the lines in \verb|STELLA|{} ought to be
absorption-dominated in order to provide better agreement with \verb|EDDINGTON|{}
and with the observed SN\,1993J. Hence, the standard value of thermalisation parameter in \verb|STELLA|{}
is $\varepsilon=1$ since 1998.
More recently, \verb|STELLA| is now the part of the latest
\verb|MESA|\footnote{Modules for Experiments in Stellar Astrophysics
\url{http://mesa.sourceforge.net/}
\citep{2011ApJS..192....3P,2013ApJS..208....4P,2015ApJS..220...15P,2018ApJS..234...34P}.}
release \citep{2018ApJS..234...34P} which is publicly available.
\verb|STELLA|{} operates under the assumption that thermalisation parameter is applied
to all species, for all transitions, and independent of electron density, which is indeed a
major simplification. However, the advantage of \verb|STELLA|{} is that it is a
hydrodynamics code, i.e., it solves implicitly coupled hydrodynamics and
multigroup radiation transport. This enables \verb|STELLA|{} to accurately capture shock propagation if the
option for artificial explosion (thermal or kinetic bomb) is set in \verb|MESA|{}.
Among other advantages is the ability
of \verb|STELLA|{} to provide reliable predictions for photometric
properties of SN explosions without requiring large computational resources.

The paper is organised as follows: We describe the method and models in
Section~\ref{sect:models}. In Section~\ref{sect:results}, we discuss our
procedure of calibration of the thermalisation parameter in \verb|STELLA|{}
using representative models of SN\,Ia, SN\,IIpec, and SN\,IIP. We
finalise the analysis in Section~\ref{sect:conclusions}, where we
specify the recommended value for the thermalisation parameter in
\verb|STELLA|{}.

\section[Input models]{Input models and Method}
\label{sect:models}

For the current study, we used three models from the literature.
The goal is to explore the impact of different degrees of
thermalisation of lines for application to:
\begin{enumerate}
\item normal SNe\,Ia --- the basic W7 model \citep{1984ApJ...286..644N}.
\item SNe\,IIpec --- the 16-7b model from \citet{2017MNRAS.469.4649M,2019MNRAS.482..438M}
exploded with the explosion energy of 2.33~foe (final kinetic energy 1.9~foe).
\item normal SNe\,IIP --- the L15 model from \citet{2000ApJS..129..625L,2017ApJ...846...37U}
exploded with the explosion energy of 1.1~foe (final kinetic energy 0.74~foe);
\end{enumerate}
These three models are not universal for the three types of SNe we discuss, but serve
as reference models for each type.

One of the main diagnostics for a possible ratio between
scattering and absorptive line opacity is the inspection of spectra, specifically how
lines redistribute energy within spectral bands. \verb|STELLA| solves
radiative transfer equations in a standard 100 frequency bins, which are not
enough to construct detailed spectra. However, the code provides spectral
energy distributions (hereafter, SED) which allow us to integrate flux in standard
BESSEL broad bands. Therefore, the
restriction on the thermalisation parameter might be determined through analysis
of the broad band magnitudes. Below we analyse separately three SN types:
SNe\,Ia, SNe\,IIpec, and SNe\,IIP.

The best way to calibrate the thermalisation parameter in \verb|STELLA|{}
is to compare to simulations calculated with advanced radiative transfer codes
that do not use the simplified treatment of line opacity. These codes
may allow photons to behave consistently with a set of calculated Non-LTE level
populations, or at least without the requirement of a free parameter governing absorption and scattering.
For SNe Type~Ia we chose the widely-used W7 model to carry out comparison to the
existing simulations done with the \verb|ARTIS|{} code \cite{2009MNRAS.398.1809K}.
Compared to \verb|STELLA|{}, \verb|ARTIS|{} has the following advantages:
\begin{enumerate}
\item Each line is treated individually, without opacity binning.
\item There is no concept of the thermalisation parameter. Radiation--matter interactions
are always treated in detail (i.e. statistical equilibrium), e.g. the full macro-atom
machinery is used to model fluorescence.
\item Although it does not use Non-LTE level populations, ARTIS calculates a Non-LTE ionisation
balance. It does this by recording detailed photoionisation rate estimators for the ground level of all ions,
and approximating photoionisation rates of excited states by assuming that they scale with the
ground level rates by the same factor as in LTE. Within an ionisation stage, the level populations are calulated from the Boltzmann distribution at the radiation temperature. \\
\end{enumerate}

On top of that, we test our calibration based on comparison to observations,
since \verb|ARTIS|{} qualitatively reproduces realistic behaviour of the
spectral energy distribution, i.e. broad band fluxes.
For SNe~IIP and SNe-IIpec, there are no well-accepted models. Therefore, we
calibrate the \verb|STELLA|{} thermalisation parameter via comparison to
observations, assuming that sophisticated codes like \verb|CMFGEN|{},
\verb|ARTIS|{}, \verb|PHOENIX|{} and others reproduce colours for observed
normal SNe~IIP and SN~1987A (i.e. SNe-IIpec) well.


\section[Results]{Discussion}
\label{sect:results}

\subsection[Application to SN\,Ia]{Application to SN~Ia}
\label{subsect:snIa}

\begin{figure*}
\centering
\includegraphics[width=0.47\textwidth]{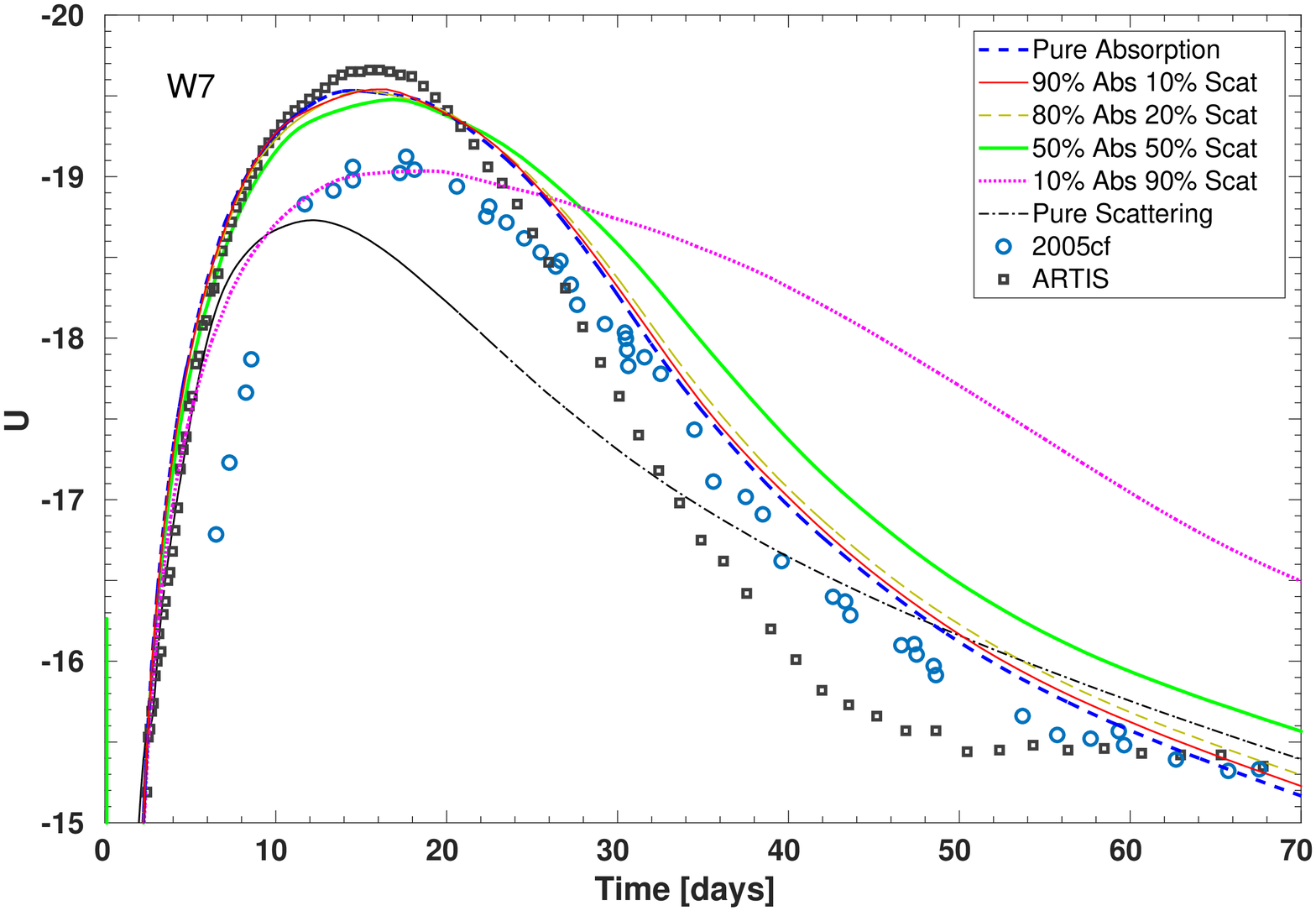}\hspace{5mm}
\includegraphics[width=0.47\textwidth]{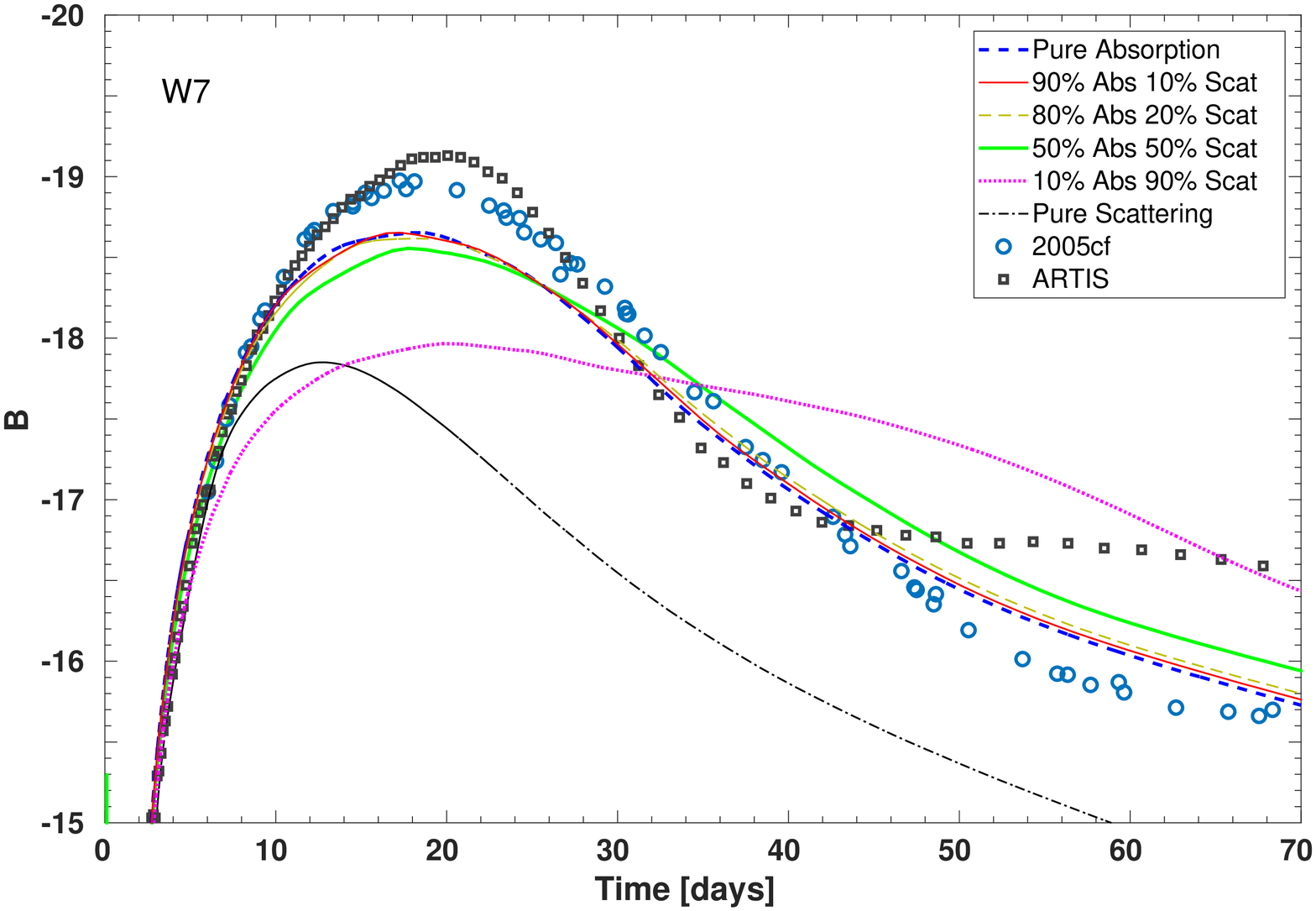}\\
\vspace{1mm}
\includegraphics[width=0.47\textwidth]{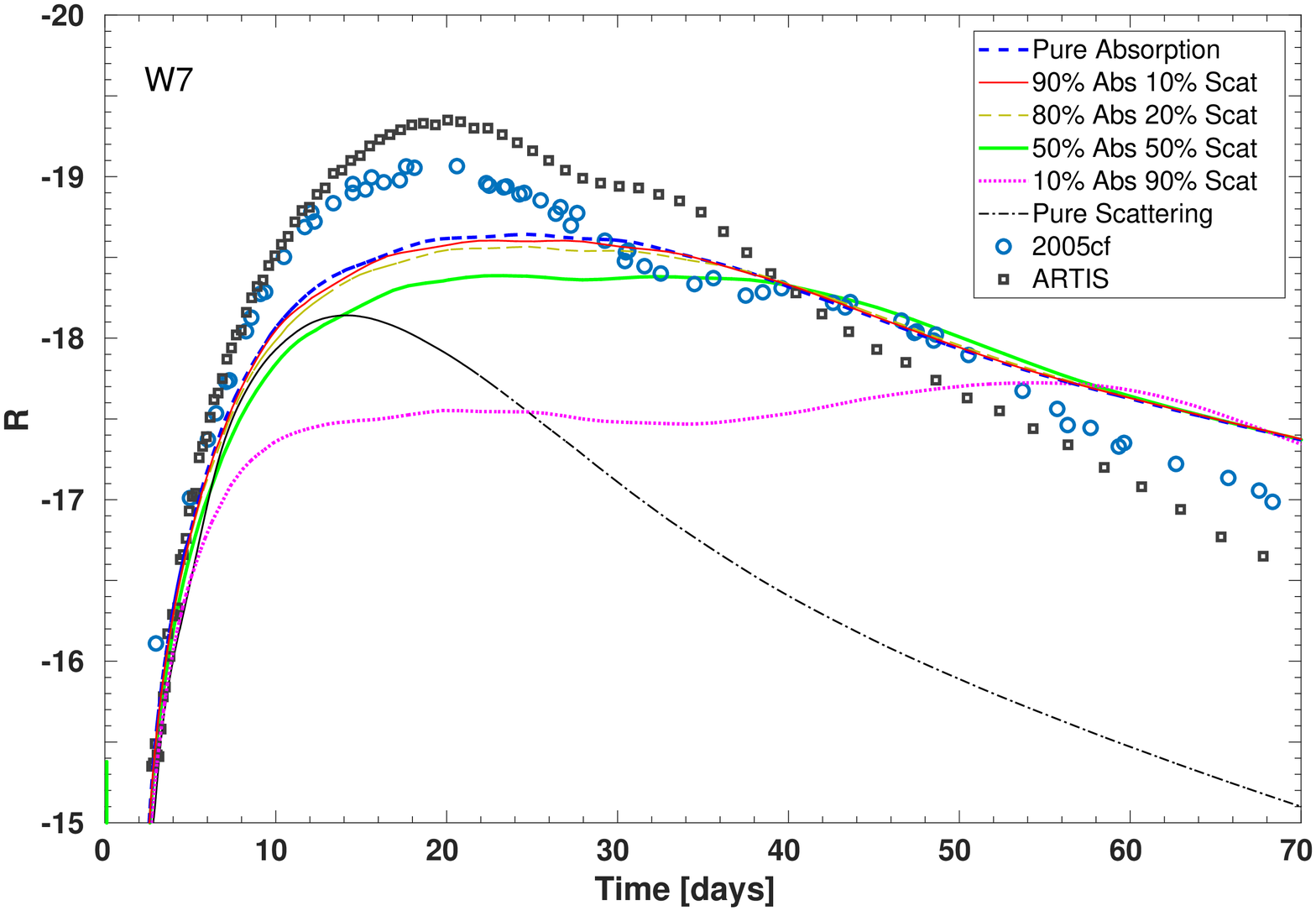}\hspace{5mm}
\includegraphics[width=0.47\textwidth]{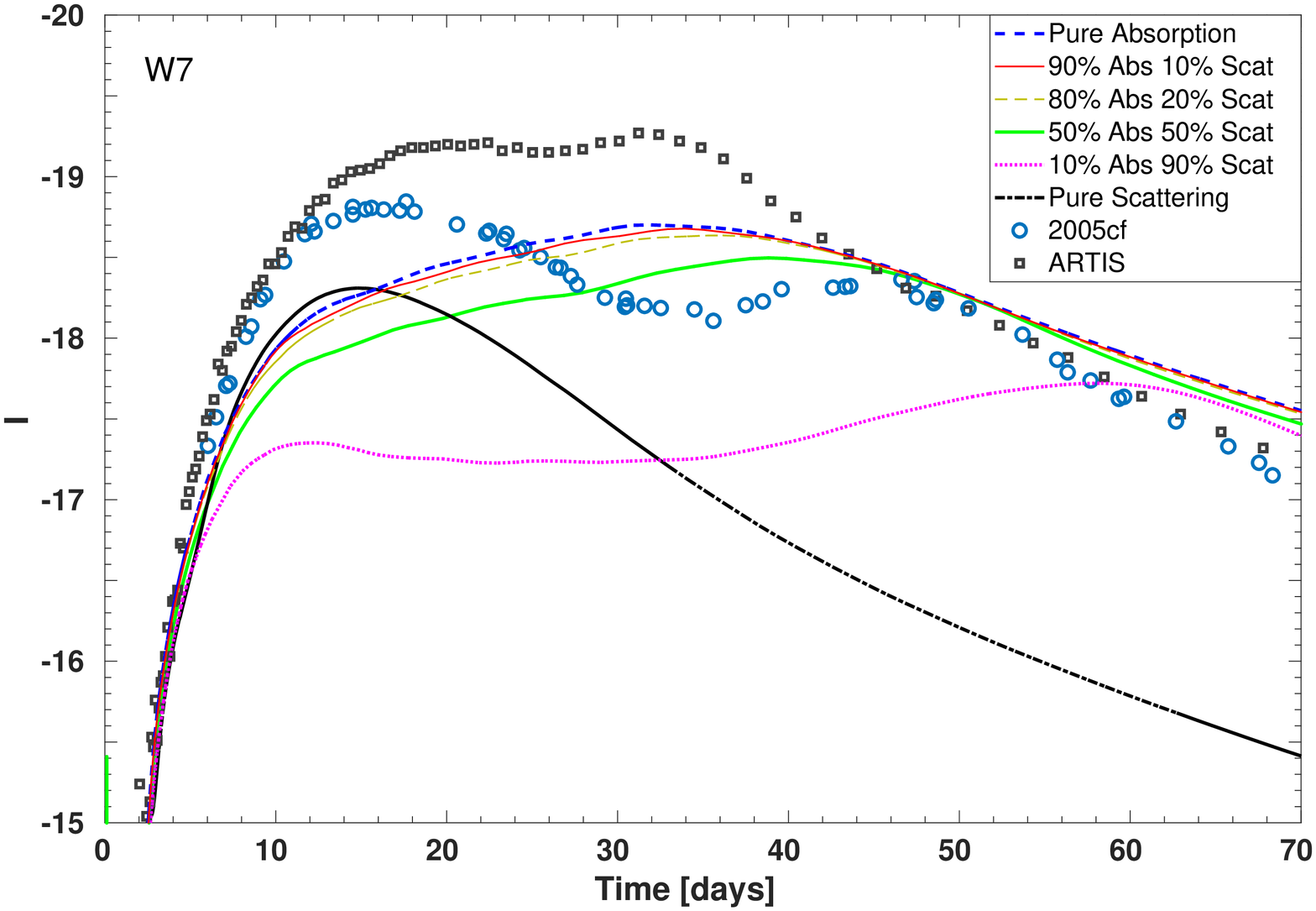}\\
\caption{The broad-band magnitudes for the W7 model
for different cases of the ratio between absorption and scattering computed
with STELLA, broad-band magnitudes for the model W7 computed with
ARTIS \citep{2009MNRAS.398.1809K},
and the observed magnitudes of SN\,2005cf \citep{2007MNRAS.376.1301P}. ``Abs''
stands for absorption fraction, and ``Scat'' stands for scattering fraction.
``Pure Absorption'' means 100\,\% of absorption, and ``Pure Scattering''
means 100\,\% of scattering.}
\label{figure:w7}
\end{figure*}

We analyse the behaviour of the numerically computed \verb|STELLA|{} light curves for comparison to
simulations done with the radiative transfer code \verb|ARTIS|{}, and observations of
the normal SN\,Ia 2005cf \citep{2007MNRAS.376.1301P}. These light curves are shown in Figure \ref{figure:w7}.

The W7 model gives a reasonable match to
the general colour evolution of SN2005cf, although there are significant discrepancies.
Even though the W7 model is one choice of many candidate explosion models, and would therefore not be expected to match all of the
features of SN\,2005cf, we still use this combination of the theoretical
model and observed data for our study. Note that we mapped into \verb|STELLA|{}
the hydrodynamical and chemical profiles of the model W7 which is exactly the same
model used for \verb|ARTIS|{} by \cite{2009MNRAS.398.1809K}.
Neither ARTIS nor STELLA simulations can explain the detailed behaviour of
SN\,2005cf, particularly, the second \emph{R} and \emph{I} maxima (see
Figure~\ref{figure:w7}).
There are a number of reasons for this. The ARTIS curves clearly demonstrate
the appearance of the second maximum which occurs a bit earlier than
the observed maximum in SN\,2005cf, and which is brighter (\emph{I}) than the observed
one. It has been shown that non-LTE (either approximate or accurate) radiative transfer may
better reproduce the second maxima \citep{2011MNRAS.417.1280B,2015MNRAS.448.2766B}.
Again, we emphasise that the model W7 itself is one particular choice of explosion model, and even theoretically-perfect radiative transfer for this model will not necessarily match observed SNe\,Ia spectra. However, a number of
explosion models can not reproduce the second maximum at all \citep{2014A&A...572A..57O}.
It could be that a slightly lower-mass model which host different thermodynamical conditions for the iron-group
elements may better reproduce the maxima occurence
\citep{2017MNRAS.470..157B,2018MNRAS.474.3931B}.
The yields of individual species like Sc, Ti, Cr, Fe, and
radioactive $^{56}$Ni{}, and their distribution in the ejecta also strongly
affect the location of the maxima \citep{2013MNRAS.429.2127B,2018ApJ...854...52S}.
As for STELLA, the code treats a limited number of
species: H, He, C, N, O, Ne, Na, Mg, Al, Si, S, Ar, Ca, stable Fe, stable Ni, stable
Co, radioactive $^{56}$Ni{}, and considers 150,000 lines in the standard settings
\citep{1995all..book.....K}. Hence line lists for individual species like Sc, Ti, Cr
are not included while they strongly contribute to line opacity.
Hence, STELLA poorly reproduces the second maxima in general, although
qualitatively represents colour evolution comparable with more sophisticated
codes even with the limited number of lines \citep{2007ApJ...662..487W}.
Certainly, new physics and extentions to the line list in
STELLA will be important areas of progress in future versions of the code.

In Figure~\ref{figure:w7}, we present the results of our simulations of the model W7
done with \verb|STELLA|{} and a range of values for the thermalisation parameter
$\varepsilon$. This allows us to explore the influence of different contributions to
absorption and scattering in bound-bound transitions.
We also superpose the results of the simulation by \citet{2009MNRAS.398.1809K}
and the observed normal SN~Ia 2005cf \citep{2007MNRAS.376.1301P}.
We show six curves computed with \verb|STELLA|{} with six corresponding
thermalisation parameters: 100\,\% absorption (the label ``Pure
Absorption''), 90\,\% absorption $+$ 10\,\% scattering (``90\% Abs
10\% Scat''), 80\,\% absorption $+$ 20\,\% scattering (``80\% Abs
20\% Scat''), 50\,\% absorption $+$ 50\,\% scattering (``50\% Abs 50\%
Scat''), 10\,\% absorption $+$ 90\,\% scattering (``10\% Abs 90\%Scat''),
and 100\,\% scattering (``Pure Scattering''). Hence, we explore the
variation of thermalisation parameter between 0 (pure scattering) and 1
(pure absorption).
The general property of the \verb|STELLA|{} light curves with different values for
$\varepsilon$ is overestimated \emph{U} magnitude, however
presumably this is an intrinsic property of the given model. While \emph{U}
magnitude is almost independent on the choice of thermalisation parameter
with moderate contribution of scattering ($\varepsilon=0.8-1$), it has a stronger
impact for \emph{R} and \emph{I}
magnitudes. $\varepsilon=0.5$ provides too broad light curve in \emph{B},
therefore, $\varepsilon=0.8-1$
is more realistic. Light curves calculated with larger contribution of
scattering ($\varepsilon<0.5$) are fully incompatible with the real observed data for
SNe\,Ia. It is worth noting that light curves with $\varepsilon=0.1$ have
two pronounced maxima which are proven observational property of SNe\,Ia. We discuss
this aspect below.

There are two major contributors to opacity in the W7 model -- iron and
calcium (considered by STELLA).
As discussed in \citet{2006ApJ...649..939K}, considering calcium lines as
purely absorptive results in overestimated \emph{I} magnitude. Therefore,
SEDONA treats calcium lines as purely scattering, while iron lines tend to
be purely absorptive. \verb|STELLA|{} applies thermalisation parameter
equally to all elements with no exceptions. With the currently implemented
line list, \verb|STELLA|{} does not resolve two maxima in \emph{I} band. Surprisingly,
the case with 90\,\% scattering does exhibit two maxima in the
\verb|STELLA|{} light curve, however both maxima are
underestimated in luminosity. This means that at least some strong lines,
i.e. most likely calcium lines, have to be treated with sufficiently low
thermalisation parameter, e.g. $\varepsilon=0.1$, i.e. almost purely scattering.
As a consequence, \emph{U} band light curve for the case of $\varepsilon=0.1$ is much closer
to the observed magnitude of SN\,2005cf. Nevertheless, the widths of the light
curves in all bands are too large and inconsistent with the observed data.

\begin{figure}
\centering
\includegraphics[width=0.5\textwidth]{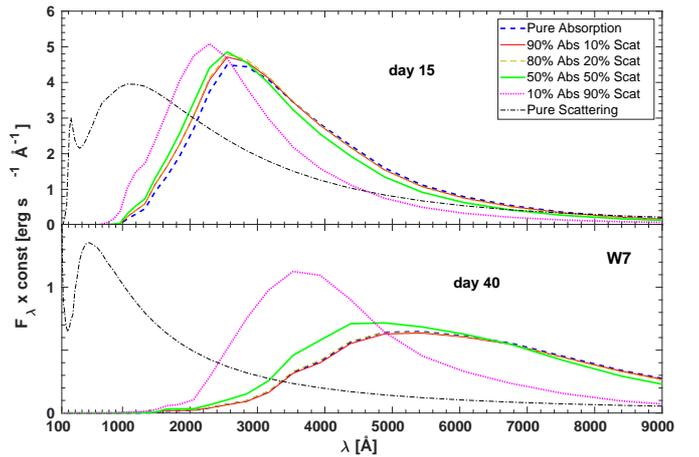}\\
\caption{Spectral energy distribution for the model W7 with different values
of thermalisation parameter: 1 (blue dashed), 0.9 (red), 0.8 (yellow
dashed), 0.5 (green), 0.1 (magenta dotted),
and 0 (black dash-dotted), at day~15 and day~40. Labels
have the same meaning as in Figure~\ref{figure:w7}.}
\label{figure:sedW7}
\end{figure}

In Figure~\ref{figure:sedW7}, we show \verb|STELLA|{} SEDs for the model
W7 for different values of thermalisation parameter at day~15 and day~40.
SED maxima for the cases with 100\,\%, 90\,\%, 80\,\%, and 50\,\%
absorption are very close to each other at day~15, although the
exact maximum wavelength differs by 300\,\AA{}. At a later epoch, day~40, the
cases of 100\,\%, 90\,\%, and 80\,\% are almost identical, while the case of 50\,\%
provides relatively bluer spectrum. Therefore, we rule out values for thermalisation
parameter below 0.8. If comparing our SEDs to the result of spectral synthesis
simulations by \verb|SEDONA|{} \citep[Fig.~5, ][]{2006ApJ...651..366K} and \verb|ARTIS|{}
\citep[Fig.~6, ][]{2009MNRAS.398.1809K}, we conclude that spectral maximum
lies around 3500\,\AA{} and around 6000\,--\,6500\,\AA{} at day~15 and
day~40, respectively, i.e. thermalisation parameter tends to be close to
unity.

We carry out a quantative analysis, particularly, we calculated the linear
Pearson's correlation coefficient with the 95\,\% confidence interval
to pick the most plausible value for the thermalisation parameter
(e.g., for the correlation between 2005cf and the STELLA curves).
We have confirmed that our findings with the correlation method are in agreement
with several alternative statistical methods to quantity the best-fit (for details, see Appendix~\ref{appendix:append}).
The observed data and the synthetic light curves are sampled on
different grids of time points.
Therefore, we first applied a 10-th order polynomial regression or
smoothing spline (using Matlab standard libraries) with the least standard
deviation criterion.
We then evaluated the resulting polynomial (or spline) on 100 equidistant time points.
We therefore converted each light curve into a vector of a uniform length, discretised onto a regular time grid.
The $\chi$--squared test evaluates statistical dependence between
standard deviations under the condition of normally distributed residuals.
$\chi$--squared test was not considered because (1) given arrays are
very limited in time, (2) the data points within a given array are not
independent, and (3) residuals do not have normal distribution.
Therefore, we choose the estimate of the correlation coefficient in the
linear regression approach as a measure of statistical coherence between vectors.
This method was applied to every curve in the study.
Correlation analysis, i.e. evaluation of statistical dependence, was based
on the calculation of the Pearson's correlation coefficient in the frame of
the linear regression model \citep{1979Afifi,1985Aivazyan}:
\begin{equation}
r=\,{\sum\limits_{i=1}^{N} (x_i-\langle x \rangle) (y_i-\langle y \rangle)
\over{\sqrt{\sigma_x \sigma_y }}} \, ,
\label{equation:corr}
\end{equation}
where $\langle x \rangle$ and $\langle y \rangle$
are the means, and $\sigma_x$ and $\sigma_y$ are the
standard deviation for light curve vectors $x$ and $y$, respectively.
The correlation analysis was carried out for each
pair of curves: U-band 2005cf and U-band STELLA curve, B-band 2005cf and B-band
STELLA curve and so on.
The interval is chosen between day~2 and day~85 for STELLA-2005cf correlation.
The same method was applied for pairs of ARTIS and STELLA curves.
The interval is 2-72 days for ARTIS-STELLA correlation.
In each pair of curves, the interval is chosen to be the longest time interval over which both curves are defined.

We analyse correlation between \verb|STELLA|{} and \verb|ARTIS|{}, and \verb|STELLA|{}
and SN\,2005cf.
The closest correlation between \verb|STELLA|{} and \verb|ARTIS|{}
is observed for the cases of
100\,\%, 90\,\%, and 80\,\% of absorption with the coefficients 0.975,
0.953, 0.980, 0.899--0.919, and 0.899--0.918 for \emph{U}, \emph{B}, \emph{V},
\emph{R}, and \emph{I}, correspondingly.
Since we look for the choice which includes at
least some scattering, we conclude that the best cases are
80\% and 90\% absorption.
We apply the same procedure to correlate \verb|STELLA|{} and SN\,2005cf.
The resulting correlation coefficients show the tight correlation for the
cases of thermalisation parameter $\varepsilon=0.5$ for \emph{U} band with coefficient 0.987,
and for $\varepsilon=0.8-1$ with coefficient
0.995, 0.999, 0.894--0.909, and 0.862-0.878 for \emph{B}, \emph{V},
\emph{R}, and \emph{I}, respectively.
To conclude, we suggest to set thermalisation parameter between 0.8 and 0.9, and
more precisely $\varepsilon=0.9$ for consistency with SNe\,II which we discuss in the sections
below.

\begin{figure*}
\centering
\includegraphics[width=0.47\textwidth]{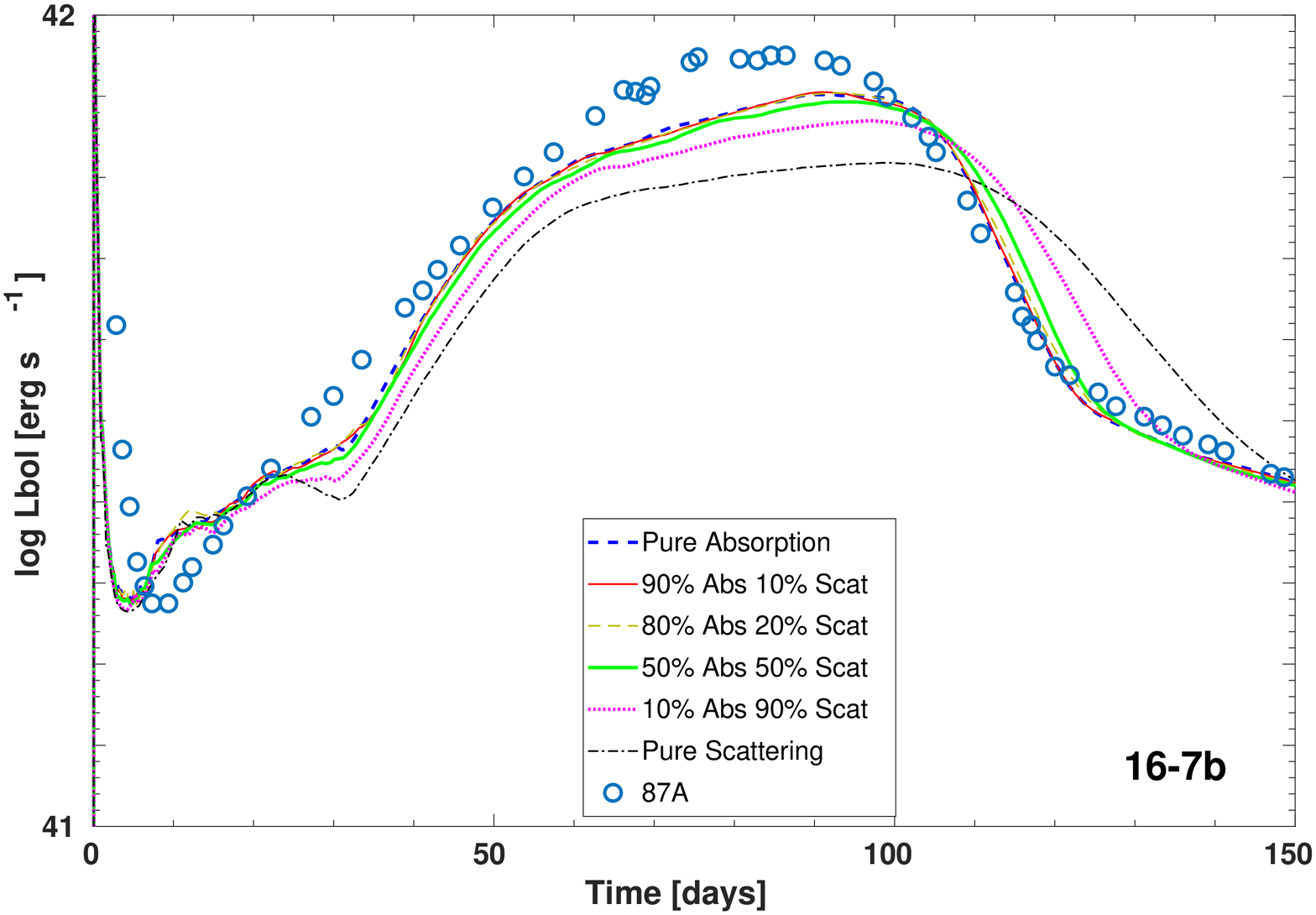}\hspace{5mm}
\includegraphics[width=0.47\textwidth]{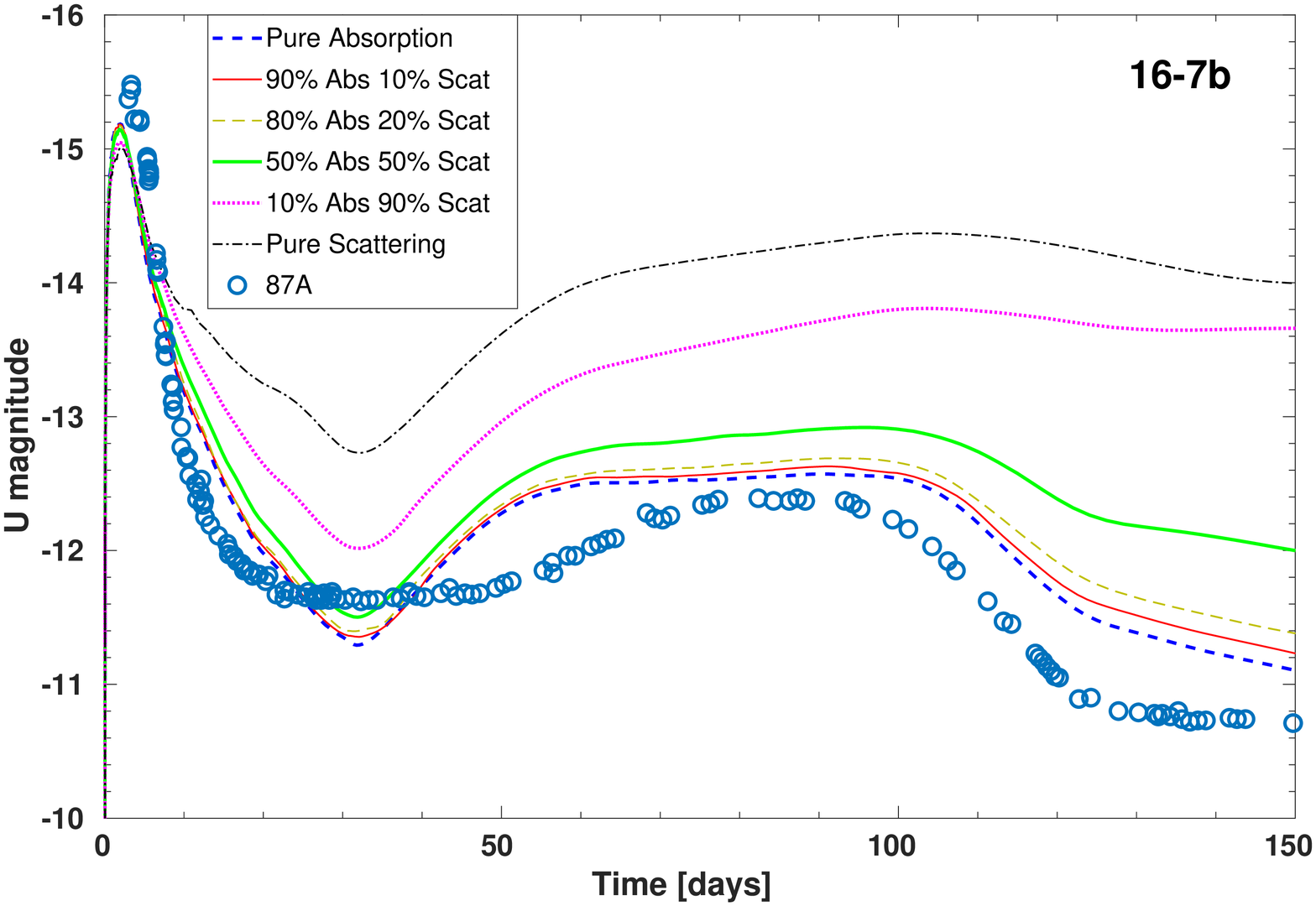}\\
\vspace{1mm}
\includegraphics[width=0.47\textwidth]{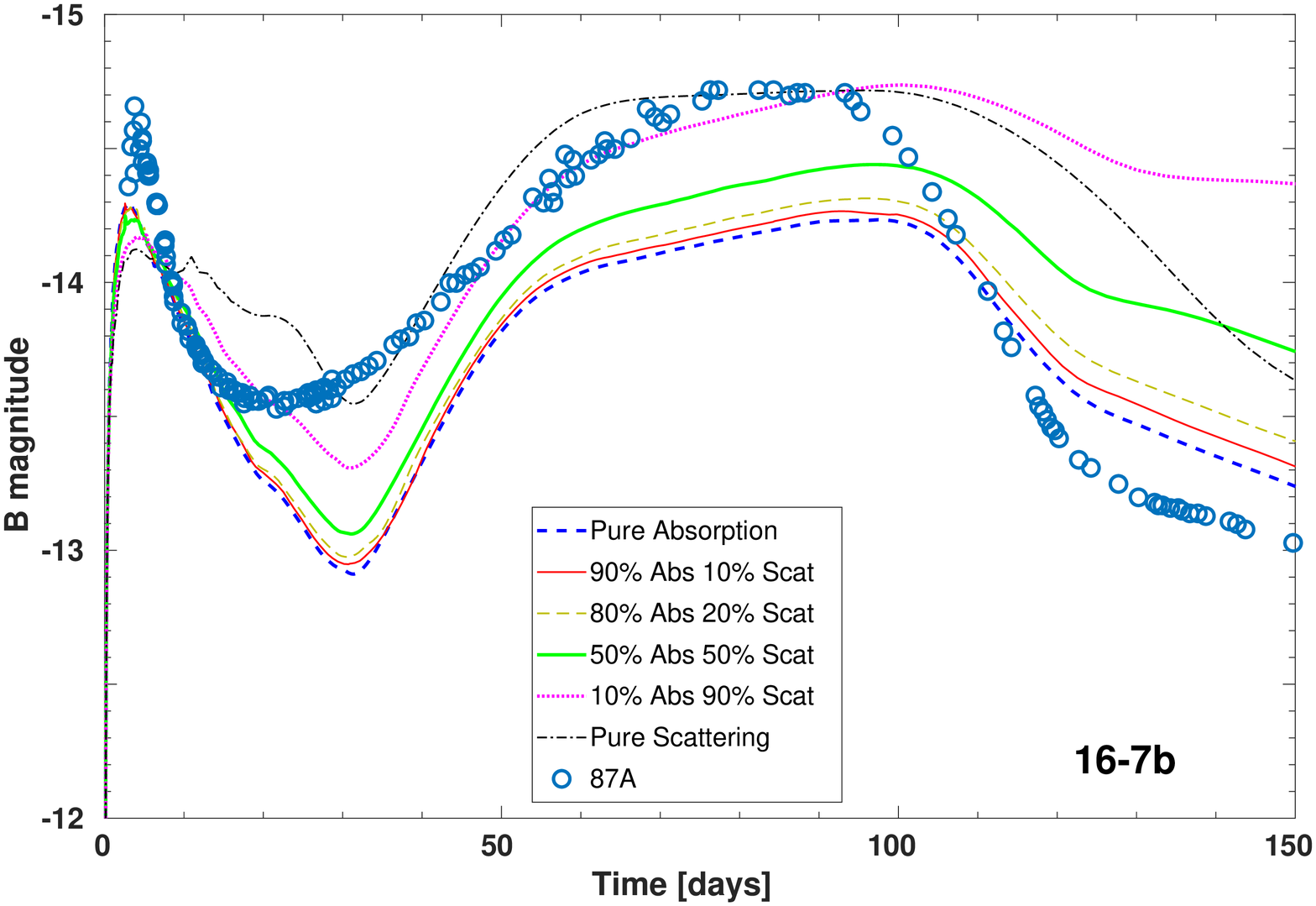}\hspace{5mm}
\includegraphics[width=0.47\textwidth]{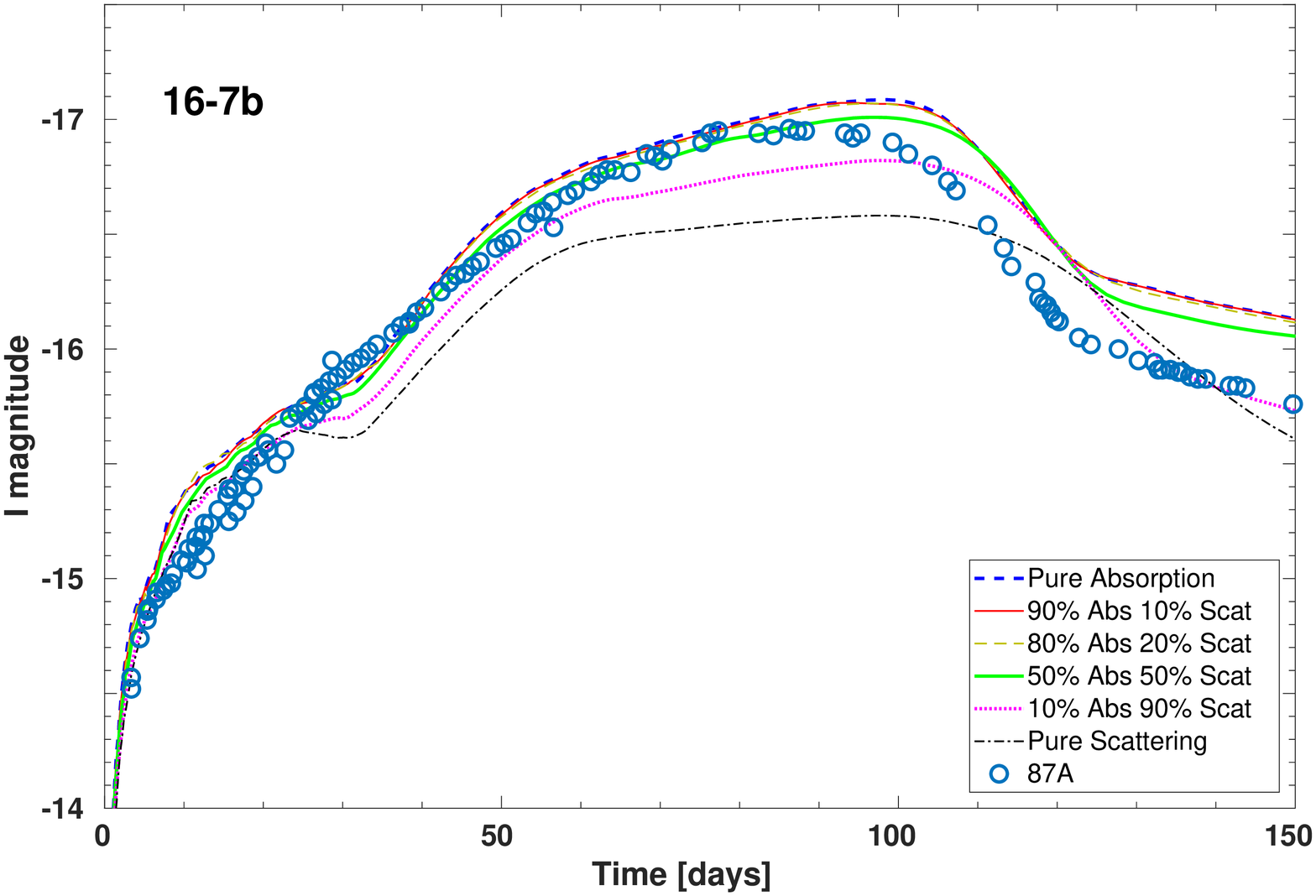}
\caption{Bolometric light curves and
\emph{U}, \emph{B} and \emph{I} broad-band magnitudes for the model 16-7b
for different cases of the ratio between absorption and scattering. Labels
have the same meaning as in Figure~\ref{figure:w7}.}
\label{figure:87A}
\end{figure*}

We conclude that correlation coefficients derived for \verb|STELLA|{}--SN\,2005cf
are the same as for \verb|STELLA|{}--\verb|ARTIS|{}. Therefore, we follow
the same procedure for SNe type IIP and 87A-like, namely we analyse
the correlation between the \verb|STELLA|{} light
curves and observed data to define the plausible thermalisation parameter.
On top of that, calibrating the code parameter on the real observations
complements our analysis, and makes future numerical simulations with \verb|STELLA|{}
even more robust.

\subsection[Application to SN\,1987A]{Application to SN~1987A}
\label{subsect:sn87a}

Modelling of the progenitor of SN~1987A is still an open question, and there
is no model that accurately reproduces all observed features.
There are single star models
\citep{1988ApJ...330..218W,1990ApJ...360..242S}, and
binary models \citep{2017MNRAS.469.4649M,2018MNRAS.473L.101U,2020ApJ...888..111O}.
Among the difficuties is the problem of producing a blue supergiant model at LMC
metallicity. Single star models are computed with artificially reduced metal
content to get a relatively compact ($\sim 50$~\Rsun{}) progenitor.
Models with the low metal content, i.e. low metallicity, provide too blue colours, particularly,
\emph{U} band magnitude is overestimated \citep{1999AstL...25..359B}.
At the same time, binary
merger models by \cite{2017MNRAS.469.4649M} have sufficient metals because
of accretion from the companion, and they are more promising as an explanation for,
particularly, the \emph{U} magnitude of SN~1987A.
Therefore, we pick up one of their recent binary models, 16-7b \citep{2019MNRAS.482..438M}, to
calibrate the thermalisation parameter in \verb|STELLA|{}.

The light curves taken with different values of thermalisation parameter are
shown in Figure~\ref{figure:87A}. Figure~\ref{figure:87A}
shows bolometric light curves (upper left) which are in acceptable
agreement with observations for the
choosen exposion energy of 2.33~foe. \emph{U} and \emph{B} magnitudes are the most affected by
variation in the thermalisation parameter, while variation in \emph{V},
\emph{R}, and \emph{I} magnitudes is not large for different parameter values.
We show \emph{U}, \emph{B}  and \emph{I}
magnitudes for demonstration. We note that we do not pay attention
to the radioactive tail, but mostly concentrate on the photospheric phase,
i.e. before approximately day~120. STELLA does not provide reliable
broad-band magnitudes after this epoch, since the SN ejecta becomes
semi-transparent, and Non-LTE treatment is requred.
For quantative analysis, we apply the same
correlation procedure as discussed in Section~\ref{subsect:snIa}. Hence we
calculate the correlation coefficients between the numerical \verb|STELLA|{} light curves in
broad bands and broad band magnitudes of SN\,1987A \citep{1987MNRAS.227P..39M}.
Light curves computed with
$\varepsilon=0.8, 0.9$ and 1 evolve close to each other. However, taking into
account the necessity of a contribution from resonant scattering, we conclude
that $\varepsilon$ has to be lower than 1, e.g. 0.8\,--\,0.9.
From the correlation analysis, the most suitable value for the thermalisation
parameter lies between 0.8 and 1, if we ignore the a priori irrelevant
values corresponding to pure scatterring and the case with 10\,\% of
absorption. The correlation coefficients are:
0.911--0.926, 0.872--0.899, 0.9--0.917, 0.936, and 0.947, for
\emph{U}, \emph{B}, \emph{V}, \emph{R}, and \emph{I}, respectively.

Additionally, we did a test using the spectral synthesis code \verb|ARTIS|
for the model B15-2 \citep{2015AandA...581A..40U} which is similar to the
model 16-7b in our study except for a zero metallicity. We run the model with a
low number of photon packets to explore the contribution by resonance
scattering and true absorption. The photon packets experience 252130, 265155,
162141, and 3152 line interactions in timesteps at day~60, day~80, day~100, and day~140,
while 4943, 5324, 3984, and 159 of those are pure scattering (i.e. no
wavelength change). Hence, the thermalisation parameter, i.e. the ratio
between the number of inelastic interactions (absorption) and total number
of interactions is: 0.98, 0.98, 0.975, and 0.95
for corresponding epochs. These numbers are very close to unity, i.e.
they show the dominant contribution of absorption to the line opacity.
However, this also demonstrates the inevitability of a non-zero scattering fraction in the line opacity.

From our analysis, we conclude that the thermalisation parameter for
hydrogen-rich SNe like SN\,1987A falls into the same interval as for
Ni-powered SNe~Ia, i.e. 0.8\,--\,0.9, with the highest plausible value of 0.9.

\subsection[Application to SN~IIP]{Application to SNe~IIP}
\label{subsect:sniip}

\begin{figure*}
\centering
\includegraphics[width=0.47\textwidth]{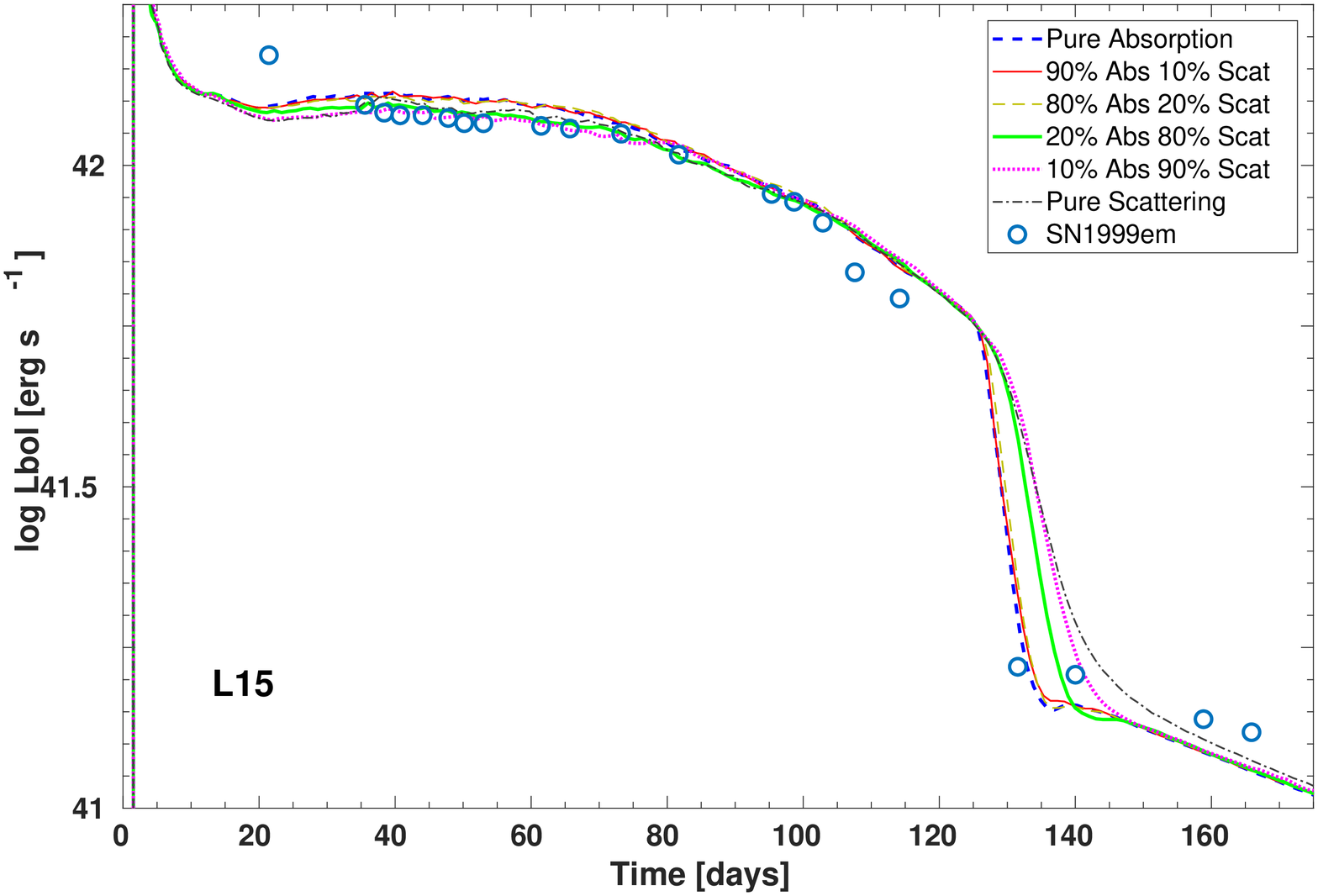}\hspace{6mm}
\includegraphics[width=0.47\textwidth]{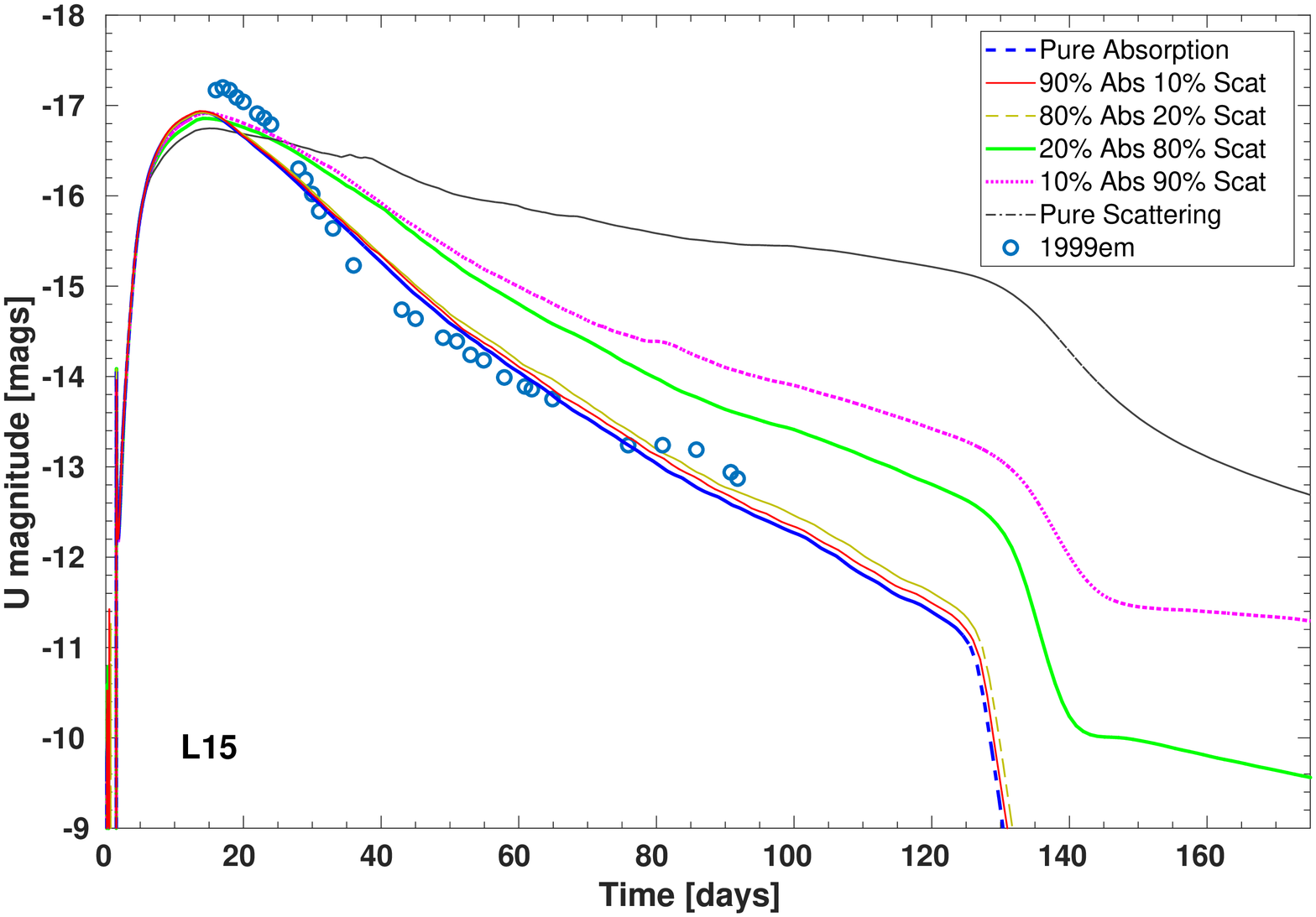}\\
\vspace{1mm}
\includegraphics[width=0.47\textwidth]{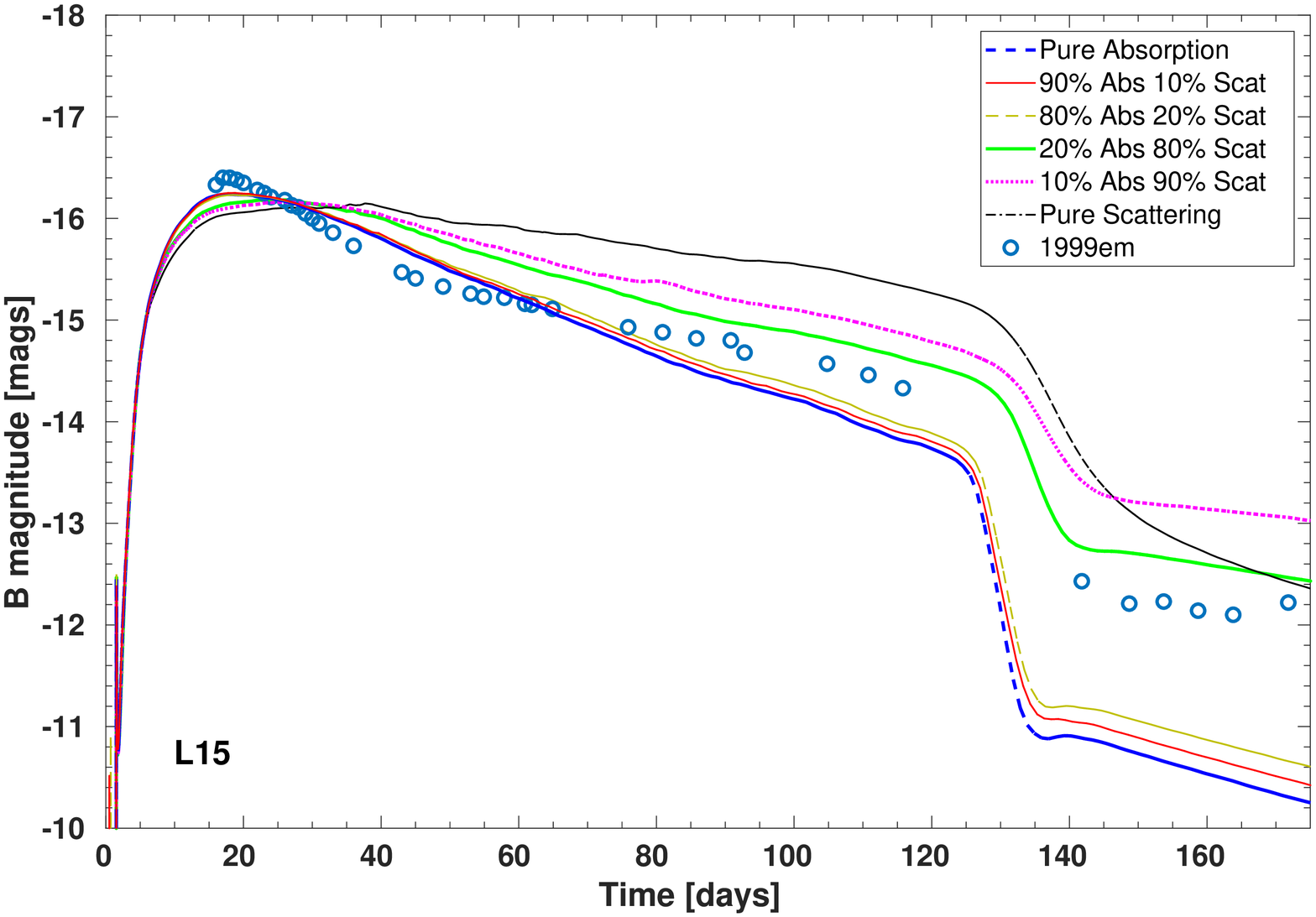}\hspace{6mm}
\includegraphics[width=0.47\textwidth]{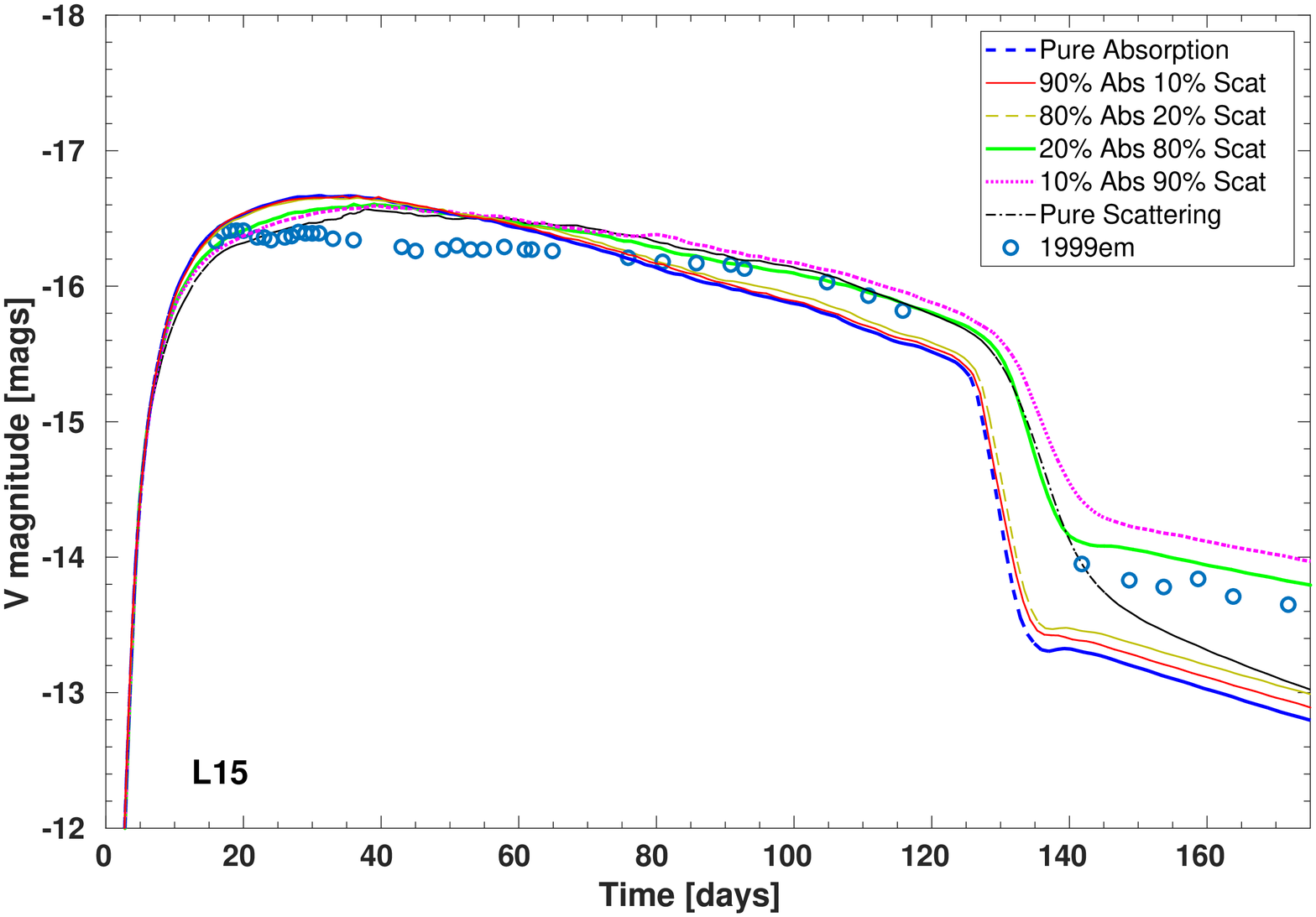}\\
\caption{Bolometric light curves and broad-band magnitudes for the model L15 for different cases of the
ratio between absorption and scattering. We superpose observed SN\,1999em as
crosses \citep{2003MNRAS.338..939E}. Labels have the same meaning as in Figure~\ref{figure:w7}
except the magenta curve, which stands now for the case of 20~\% of
absorption and 80~\% of scattering.}
\label{figure:lsc15nbands}
\end{figure*}

The model L15 is the accepted model which reproduces bolometric properties
of SN\,1999em \citep{2003MNRAS.338..939E,2007A&A...461..233U,2017ApJ...846...37U}.
In Figure~\ref{figure:lsc15nbands}, we show bolometric light curves and broad-band
magnitudes for the model L15 for different cases of the
ratio between absorption and scattering. Firstly, we present bolometric
light curves for the subset of the model L15 (the upper left plot in Figure~\ref{figure:lsc15nbands})
to demonstrate that the model is indeed sufficient to explain the bolometric
luminosity of SN\,1999em. As in previous sections, we vary the thermalisation
parameter between 0 and 1 for the model L15.
The overall differences between bolometric light curves are not large. There
is a tiny variation in early plateau luminosity (maximum 0.05~dex), while
there is a noticable difference in the behaviour of the transition of the light
curve to the radioactive tail. Nevertheless, the treatment of lines is indeed not
important for the bolometric properties at earlier epoch because lines contribute
less significantly than the
continuum opacity during the early phase and especially during the
relaxation after shock breakout. However, the lines start
playing a significant role at day~20 for our model L15 (see Figure~\ref{figure:lsc15nbands}).
In any case, the bolometric light curve is rather insensitive to different
degrees of thermalisation in lines because it reflects the overall energy budget
(which is conserved), while different thermalisation values only cause
redistribution of energy between parts of the spectrum.

\begin{figure}
\centering
\includegraphics[width=0.5\textwidth]{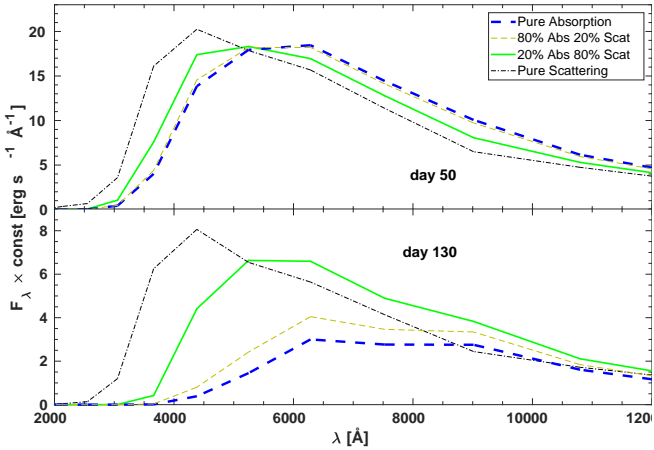}
\caption{Spectral energy distribution for the model L15 for cases of 0~\%
(blue dashed), 20~\% (red), 80~\% (green dashed),
and 100~\% (magenta) of scattering at day~50 and day~130. Labels have the same meaning as
in Figure~\ref{figure:lsc15nbands}.}
\label{figure:sed99em}
\end{figure}

\begin{figure}
\centering
\includegraphics[width=0.5\textwidth]{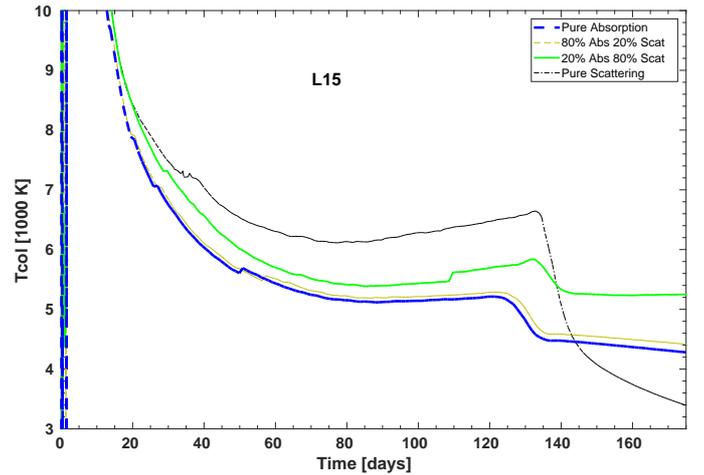}
\caption{Colour temperature evolution for the model L15 for cases of 0~\%
(blue dashed), 20~\% (red), 80~\% (green dashed), and 100~\% (magenta) of scattering.
Labels have the same meaning as in Figure~\ref{figure:lsc15nbands}.}
\label{figure:Tcol}
\end{figure}

With the pure scattering line
opacity, \emph{U} and \emph{B} remain blue for the entire plateau.
Blue photons from the decay of \Ni{} are accumulated in the region where they
are born. Line opacity is much higher for blue photons than for
redder photons, with the effect that blue photons require a longer time to diffuse through the optically-thick medium.
Therefore, scattering-dominated line opacity provides a larger fraction of
blue photons remaining in the inner region of the SN ejecta.
Assuming SN~1999em to be a normal SN~IIP, \emph{U} fades during
the plateau phase with the slope 4--5~mags during 100~days, and \emph{B} magnitude
drops with the slope 2~mags during this period. This is consistent with our
modelled curves with scattering fraction of 0--20~\%{}.
Observations by \citet{2014MNRAS.442..844F},
\citet{2016MNRAS.459.3939V} and \citet{2019arXiv190309048S}
also confirm these slopes and present a large set of SNe~IIP which
show standard decline of about 2~mags in \emph{B} band
during 100~days. \emph{R} and \emph{I} light curves are not strongly
affected by the value of the thermalisation parameter.
We conclude that introducing a larger fraction of scattering to the line opacity leads
to long-lasting blue colours which is not supported by observations of normal SNe~IIP.

We follow the same correlation procedure as we applied for SNe~Ia and
SNe~1987A-like. We found that the light curves with $\varepsilon=0.8$, 0.9,
and 1 are the most correlated with the observed broad-band magnitudes of SN~1999em.
The correlation coefficients are: 0.98, 0.992, and 0.992 for \emph{U},
\emph{B}, and \emph{V}, respectively.

In Figure~\ref{figure:sed99em}, we show spectral energy distributions for the model
L15 for cases of 0~\%, 20~\%, 80~\%,
and 100~\% of scattering at day~50 and day~130. There is an extra
blue flux for those cases where scattering fraction is higher, which leads
to higher colour temperature. Figure~\ref{figure:Tcol} shows corresponding
colour temperature evolution for these cases. While pure absorption (what means
0~\% scattering) and 20~\% scattering cases are suitable for normal SNe~IIP
\citep[see e.g., ][]{2009ApJ...701..200B},
curves with larger scattering contribution are totally unrealistic, because
of relatively too-blue colour at the end of their plateau.

To conclude, we find the value $\varepsilon=0.9$ provides the best match to
colours in normal SN\,IIP.

\section[Conclusions]{Conclusions}
\label{sect:conclusions}

We explored the treatment of lines in the
hydrodynamics radiative transfer code \verb|STELLA|{}. Previously, all lines
in the code were considered as purely absorptive, i.e. a photon was
immediately thermalised on interaction with matter. We analysed the
impact of introduction scattering into the line treatment using three reference models,
W7 \citep{1984ApJ...286..644N},
16-7b \citep{2017MNRAS.469.4649M,2019MNRAS.482..438M}, and
L15 \citep{2000ApJS..129..625L,2017ApJ...846...37U},
to illustrate the impact of different degree of
thermalisation in lines on the behaviour of light curves and SEDs of normal
SNe\,Ia, SNe\,IIpec,
and normal SNe\,IIP. We analysed light curves in the broad bands in the context of
the more sophisticated simulations done with the code \verb|ARTIS|{}, and
well-observed SN\,2005cf, SN\,1987A, and SN\,1999em.

We found that the most suitable value for the thermalisation parameter
$\varepsilon$, i.e. the relative contribution of absorption to overall line
opacity, lies between 0.8 and 0.9 for the three types of SNe considered.
The scattering due to lines should be less than 10--20\,\% to prevent blue flux from exceeding observed SNe.

Our recommendation is to use $\varepsilon=0.9$ in all future simulations
that will be done with the code \verb|STELLA|{}
which is a part of the latest \verb|MESA|{} release \citep{2018ApJS..234...34P}.

\section*{Acknowledgments}
We thank the referee Prof. Saurabh W. Jha for his interest in our study and careful analysis of
our methodology and results that improved the paper.
AK is supported by the Alexander von Humboldt Foundation.
LS acknowledges support from STFC through grant, ST/P000312/1.
SB and PB are sponsored by grant RSF\,18-12-00522. 
Some of this work was performed using the Cambridge Service for Data Driven
Discovery (CSD3), part of which is operated by the University of Cambridge
Research Computing on behalf of the STFC DiRAC HPC Facility
(www.dirac.ac.uk).  The DiRAC component of CSD3 was funded by BEIS capital
funding via STFC capital grants ST/P002307/1 and ST/R002452/1 and STFC
operations grant ST/R00689X/1.  DiRAC is part of the National
e-Infrastructure.
The authors thank Viktor Utrobin, Thomas Janka for discussions which
initiated this study, and
Markus Kromer and Stuart Sim for clarifications about \verb|ARTIS|{}
simulations, Anders Jerkstrand, Ildar Khabibullin and Marat Potashov for overall discussions.

\addcontentsline{toc}{section}{Acknowledgments}

\section*{Data availability}

The data computed and analysed for the current study are available via link
{\tiny{\url{https://wwwmpa.mpa-garching.mpg.de/ccsnarchive/data/Kozyreva2018/index.html}.}}

\bibliographystyle{mnras}
\bibliography{references}
\appendix
\section[Additional statistics]{Additional statistics}
\label{appendix:append}

We carried out additional statistical tests to verify the results of the
correlation method that we used in Section~\ref{sect:results}.

\begin{figure*}
\centering
\includegraphics[width=0.3\textwidth]{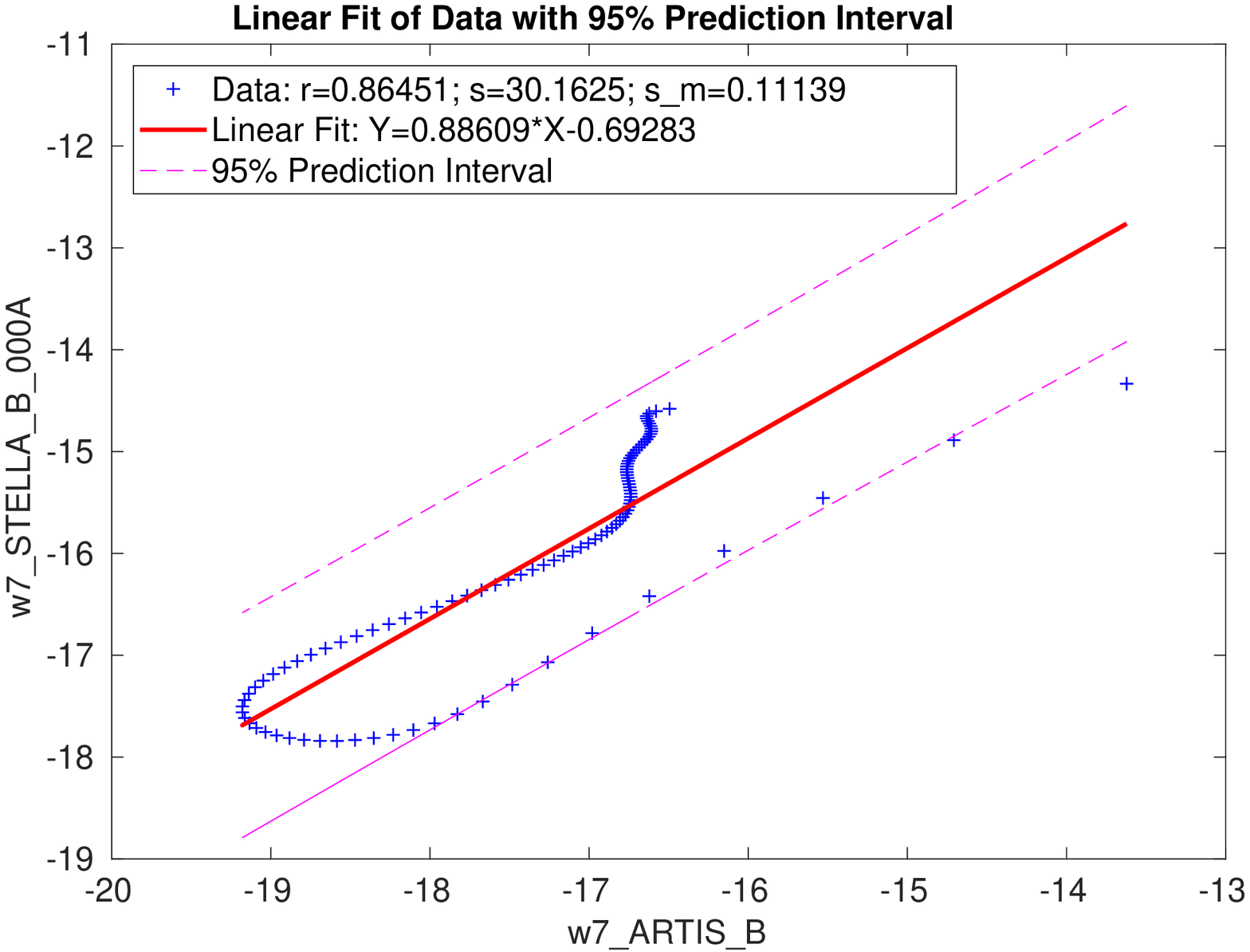}\hspace{5mm}
\includegraphics[width=0.3\textwidth]{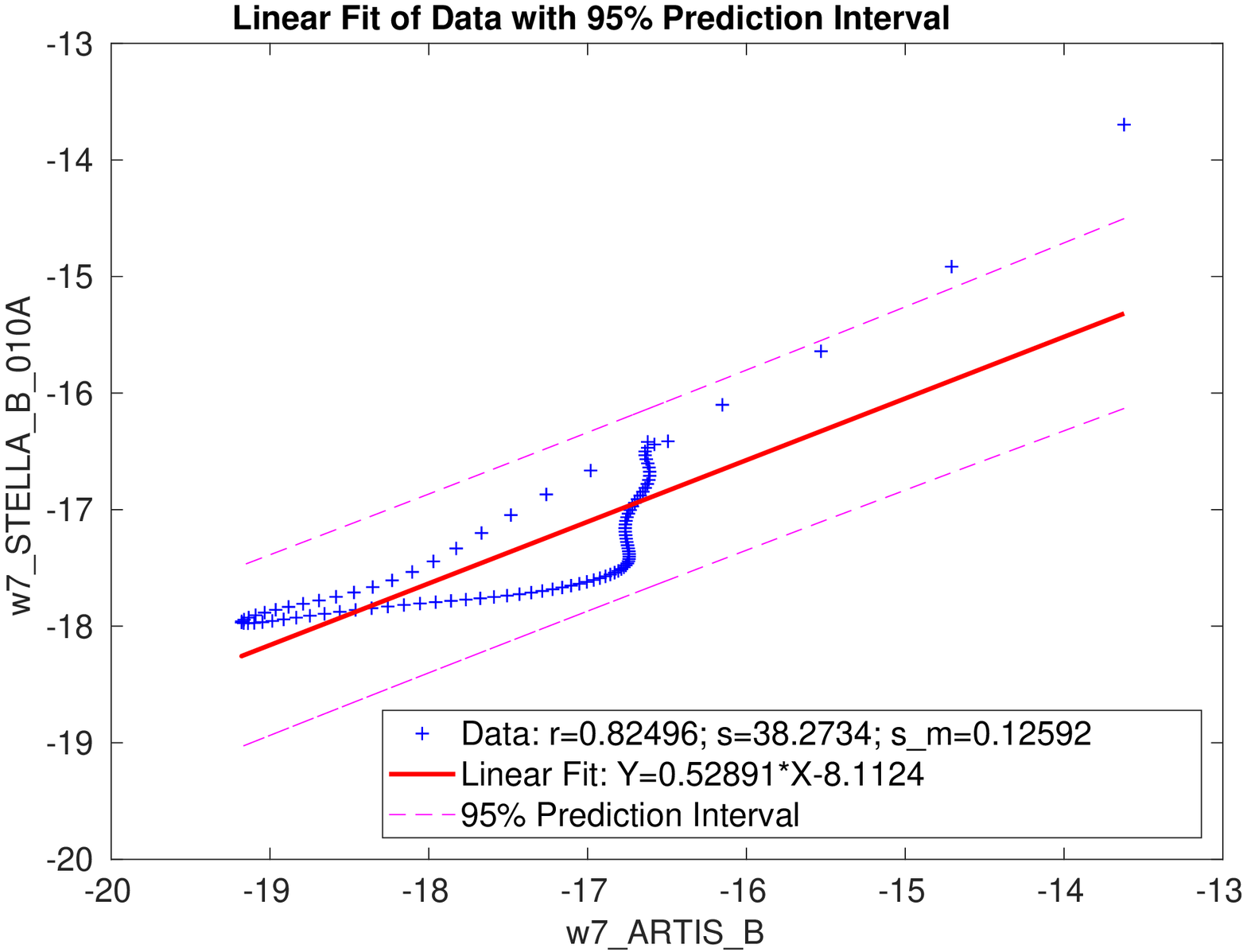}\hspace{5mm}
\includegraphics[width=0.3\textwidth]{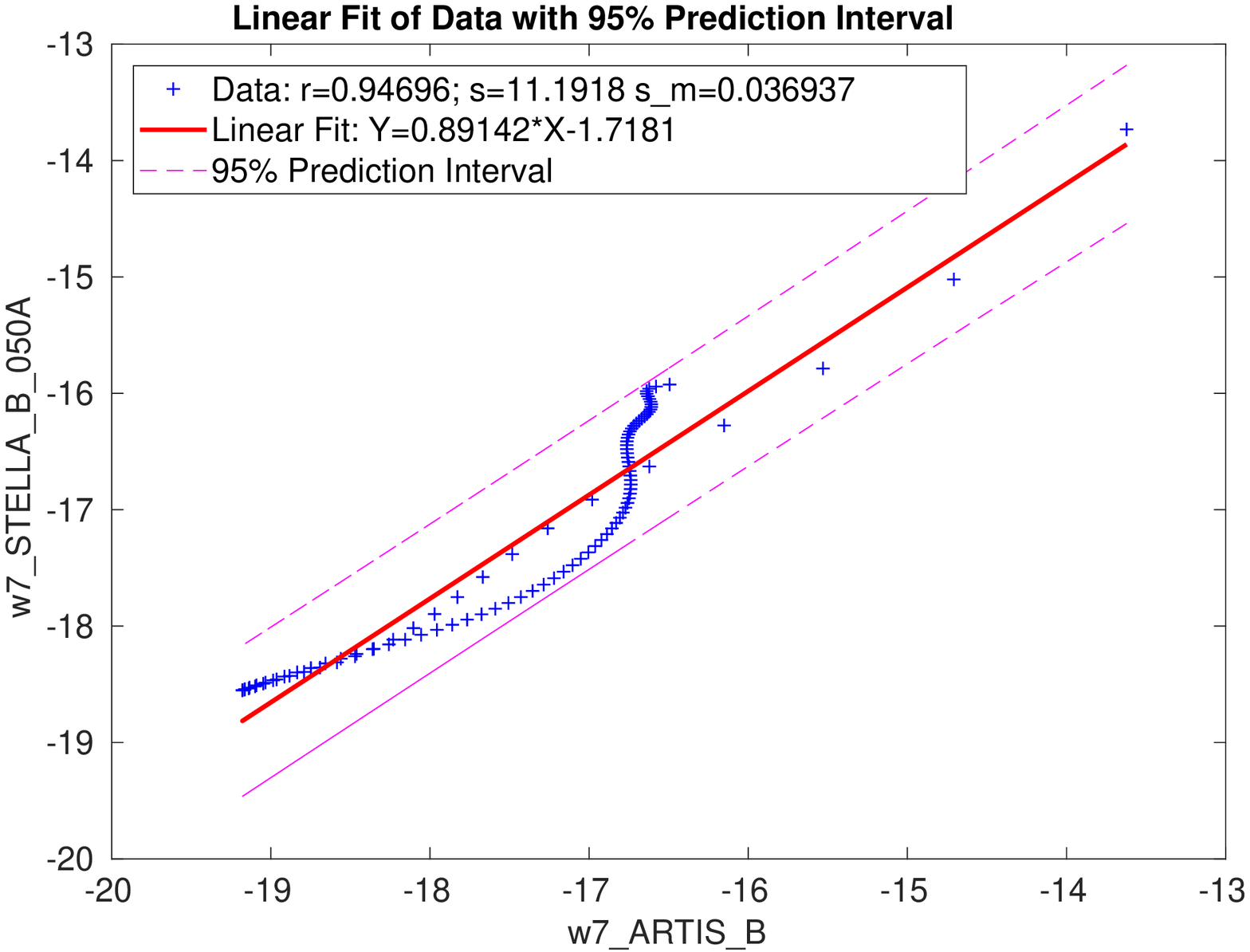}\\
\vspace{1mm}
\includegraphics[width=0.3\textwidth]{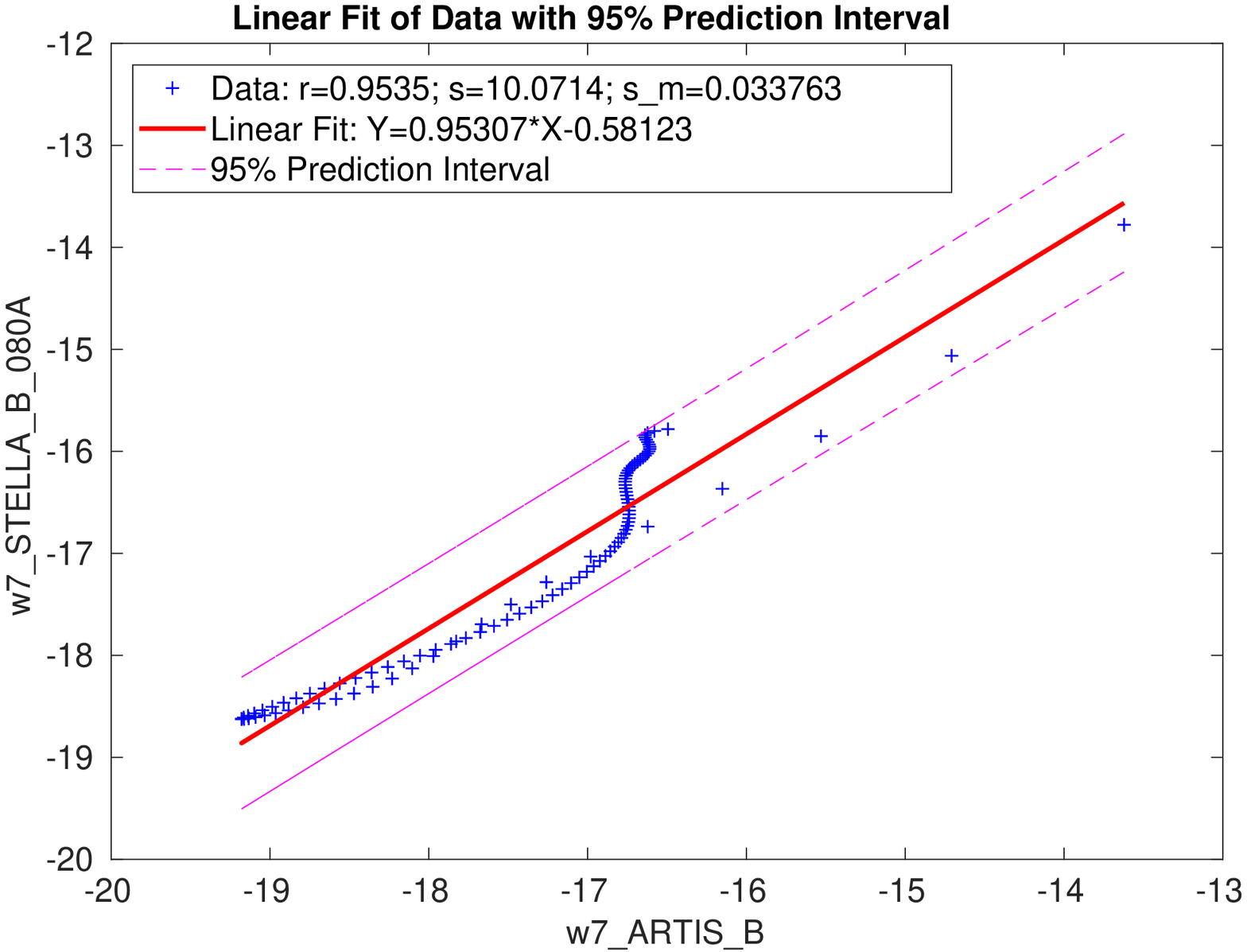}\hspace{5mm}
\includegraphics[width=0.3\textwidth]{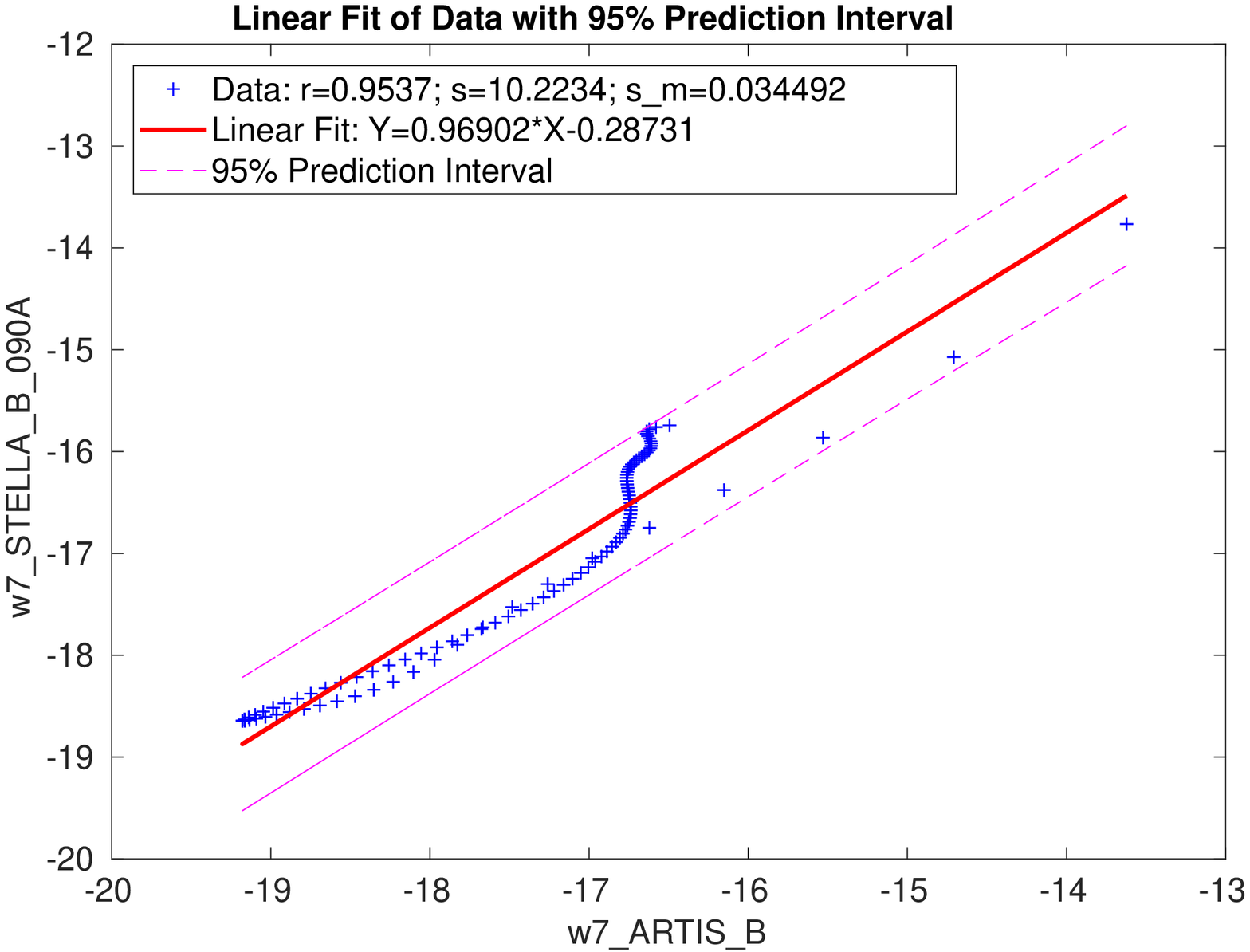}\hspace{5mm}
\includegraphics[width=0.3\textwidth]{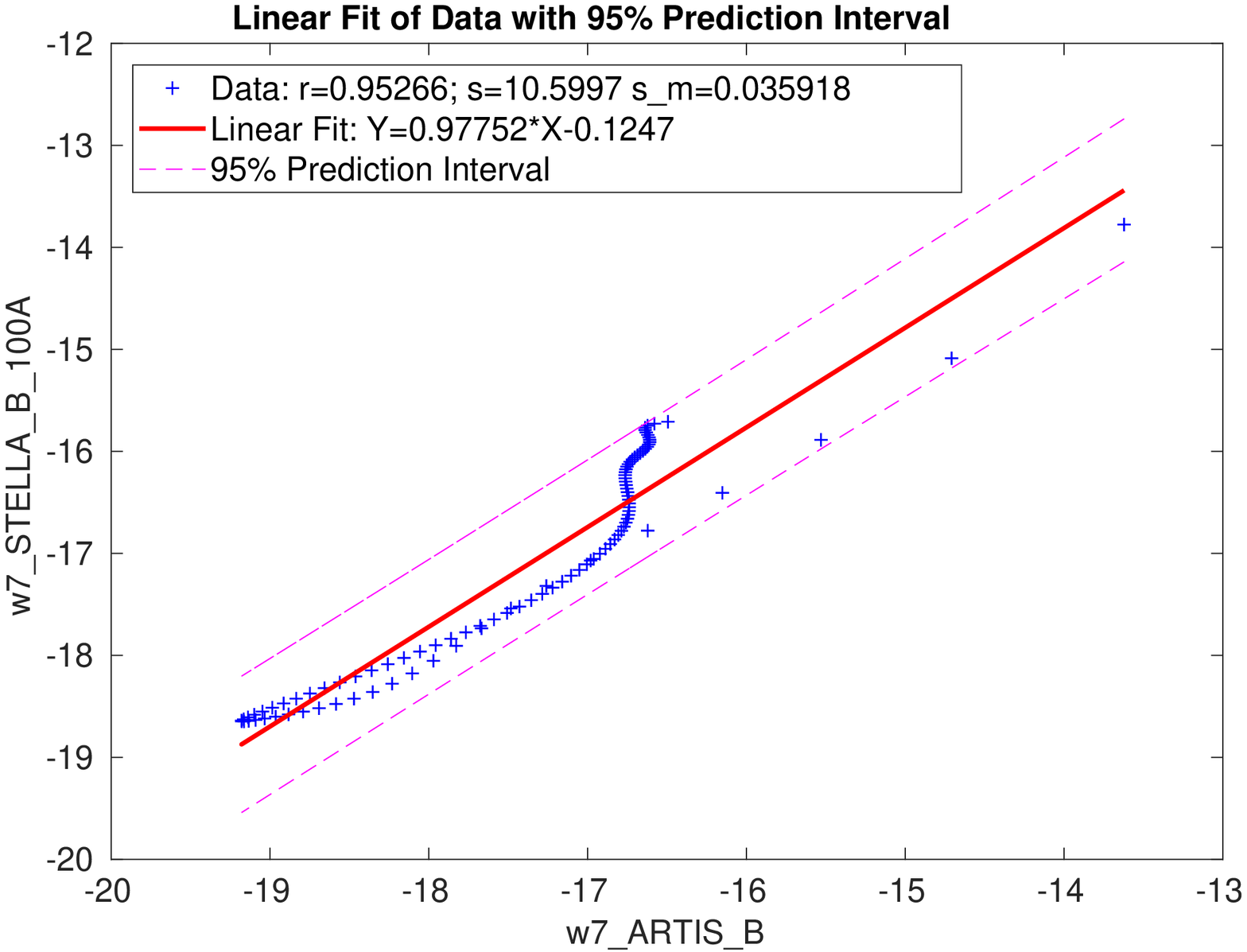}\\
\caption{The \emph{B}-band magnitudes of the W7 model
computed with STELLA using different assumed ratios between absorption and scattering, and the W7 model computed with ARTIS \citep{2009MNRAS.398.1809K}.}
\label{figure:w7corr}
\end{figure*}

Let us consider the pairs of arrays from our study, namely, the \emph{B}-band magnitudes calculated with
\verb|ARTIS| and \verb|STELLA| (with different degree of thermalisation).
In Figure~\ref{figure:w7corr}, we demonstrate the pairs of arrays:
\verb|ARTIS| \emph{B}-band magnitude versus \verb|STELLA| \emph{B}-band magnitude computed with
different value of the thermalisation parameter.
We calculated the linear regression coefficient for each pair, assuming
$Y=A\times X + B$, where $X$ is \verb|ARTIS| magnitude and $Y$ is \verb|STELLA|
magnitude.

We consider the statistic:

\begin{equation}
s = \sum\limits_{i} \left[(y_i - \langle y \rangle) -
(x_i - \langle x \rangle)\right]^{\,2}
\label{equation:SWJ}
\end{equation}

and slightly modified statistic:

\begin{equation}
s_m = \, \sum\limits_{i} \left[{{y_i - \langle y \rangle}\over{\langle y \rangle}}
- {{x_i - \langle x \rangle}\over {\langle x \rangle}} \right]^{\,2} \, .
\label{equation:mSWJ}
\end{equation}
The minimal value for these statistics will indicate the best match
between arrays $X$ and $Y$.

In Table~\ref{table:coef1}, we list the resulting coefficients. Here, $A$ is
linear regression coefficient, $r$ is linear correlation coefficient (see
Equation~\ref{equation:corr} in Section~\ref{subsect:snIa}), and two coefficients
$s$ and $s_m$ counting for the proposed additional statistics.
The maximal correlation coefficient correspond to minimal
$s$ and $s_m$ coefficients. We note though that maximal $r$ and minimal
$s/s_m$ do not necessarily correspond to the maximal linear regression coefficient (its closure to 1).
We highlight in bold face the best match according to all considered statistics.
Hence, we rely on the maximal correlation coefficient in the main body of
the paper as the diagnostic for the best match between light curves and
the selection of the best value for the thermalisation parameter. We list the
coefficients for the combination \verb|ARTIS|--\verb|STELLA| in the
\emph{U}, \emph{V}, \emph{R}, and \emph{I} broad bands in
Table~\ref{table:coef2}.

\begin{table*}
\caption{The linear regression coefficient $A$, linear correlation
coefficient $r$, and the statistics $s$ and $s_m$ for the pairs of light curves
in \emph{B} broad band computed with STELLA and ARTIS. The line in bold face
corresponds to the maximum correlation coefficient and minimum $s$ and $s_m$
coefficients.}
\label{table:coef1}
\begin{tabular}{l|l|l|l|l|l}
    &          & $A$ & $r$ & $s$  & $s_m$\\
\hline
B\underline{\,}ARTIS & B\underline{\,}STELLA\underline{\,}000A & 0.88609 & 0.86451 & 30.1625 & 0.11139\\
B\underline{\,}ARTIS & B\underline{\,}STELLA\underline{\,}010A & 0.52891 & 0.82496 & 38.2734 & 0.12592\\
B\underline{\,}ARTIS & B\underline{\,}STELLA\underline{\,}050A & 0.89142 & 0.94696 & 11.1918 & 0.036937\\
B\underline{\,}ARTIS & B\underline{\,}STELLA\underline{\,}080A & 0.95307 & 0.95350 & 10.0714 & 0.033763\\
\textbf{B\underline{\,}ARTIS} & \textbf{B\underline{\,}STELLA\underline{\,}090A} &\textbf{0.96902} &\textbf{0.95370} &\textbf{10.2234} &\textbf{0.034492}\\
B\underline{\,}ARTIS & B\underline{\,}STELLA\underline{\,}100A & 0.97752 & 0.95266 & 10.5997 & 0.035918\\
\end{tabular}
\end{table*}

\begin{table*}
\caption{The same as in Table~\ref{table:coef1} for the W7 U,V,R,I-band STELLA magnitudes and ARTIS magnitudes.}
\label{table:coef2}
\begin{tabular}{l|l|l|l|l|l}
       &       & $A$ & $r$ & $s$  & $s_m$\\
\hline
U\underline{\,}ARTIS & U\underline{\,}STELLA\underline{\,}000A & 0.64213 & 0.97445 & 40.67760 & 0.13968\\
U\underline{\,}ARTIS & U\underline{\,}STELLA\underline{\,}010A & 0.44048 & 0.80076 & 114.35609 & 0.41186\\
U\underline{\,}ARTIS & U\underline{\,}STELLA\underline{\,}050A & 0.81061 & 0.95934 & 25.15613 & 0.09365\\
U\underline{\,}ARTIS & U\underline{\,}STELLA\underline{\,}080A & 0.88321 & 0.97406 & 15.11910 & 0.05550\\
U\underline{\,}ARTIS & U\underline{\,}STELLA\underline{\,}090A & 0.90048 & 0.97608 & 13.59926 & 0.04947\\
U\underline{\,}ARTIS & U\underline{\,}STELLA\underline{\,}100A & 0.91200 & 0.97707 & 12.81158 & 0.04621\\
\hline
V\underline{\,}ARTIS & V\underline{\,}STELLA\underline{\,}000A & 0.88609 & 0.86451 & 30.1625 & 0.11139\\
V\underline{\,}ARTIS & V\underline{\,}STELLA\underline{\,}010A & 0.52891 & 0.82496 & 38.2734 & 0.12592\\
V\underline{\,}ARTIS & V\underline{\,}STELLA\underline{\,}050A & 0.89142 & 0.94696 & 11.1918 & 0.036937\\
V\underline{\,}ARTIS & V\underline{\,}STELLA\underline{\,}080A & 0.95307 & 0.95350 & 10.0714 & 0.033763\\
V\underline{\,}ARTIS & V\underline{\,}STELLA\underline{\,}090A & 0.96902 & 0.95370 & 10.2234 & 0.034492\\
V\underline{\,}ARTIS & V\underline{\,}STELLA\underline{\,}100A & 0.97752 & 0.95266 & 10.5997 & 0.035918\\
\hline
R\underline{\,}ARTIS & R\underline{\,}STELLA\underline{\,}100A & 0.84774 & 0.86779 & 27.72292 & 0.09397\\
R\underline{\,}ARTIS & R\underline{\,}STELLA\underline{\,}100A & 0.25819 & 0.51068 & 79.17357 & 0.24369\\
R\underline{\,}ARTIS & R\underline{\,}STELLA\underline{\,}100A & 0.55486 & 0.84413 & 34.52384 & 0.10564\\
R\underline{\,}ARTIS & R\underline{\,}STELLA\underline{\,}100A & 0.61134 & 0.89904 & 25.67382 & 0.07864\\
R\underline{\,}ARTIS & R\underline{\,}STELLA\underline{\,}100A & 0.62611 & 0.91026 & 23.65941 & 0.07251\\
R\underline{\,}ARTIS & R\underline{\,}STELLA\underline{\,}100A & 0.63532 & 0.91865 & 22.23886 & 0.06819\\
\hline
I\underline{\,}ARTIS & I\underline{\,}STELLA\underline{\,}100A & 0.84681 & 0.78094 & 39.04812 & 0.13263\\
I\underline{\,}ARTIS & I\underline{\,}STELLA\underline{\,}100A & 0.26291 & 0.47073 & 63.65931 & 0.19110\\
I\underline{\,}ARTIS & I\underline{\,}STELLA\underline{\,}100A & 0.63093 & 0.84502 & 23.93898 & 0.07053\\
I\underline{\,}ARTIS & I\underline{\,}STELLA\underline{\,}100A & 0.69274 & 0.89938 & 16.82663 & 0.04937\\
I\underline{\,}ARTIS & I\underline{\,}STELLA\underline{\,}100A & 0.70765 & 0.91048 & 15.28726 & 0.04483\\
I\underline{\,}ARTIS & I\underline{\,}STELLA\underline{\,}100A & 0.71676 & 0.91804 & 14.25680 & 0.04180\\
\end{tabular}
\end{table*}

\begin{figure*}
\centering
\includegraphics[width=0.3\textwidth]{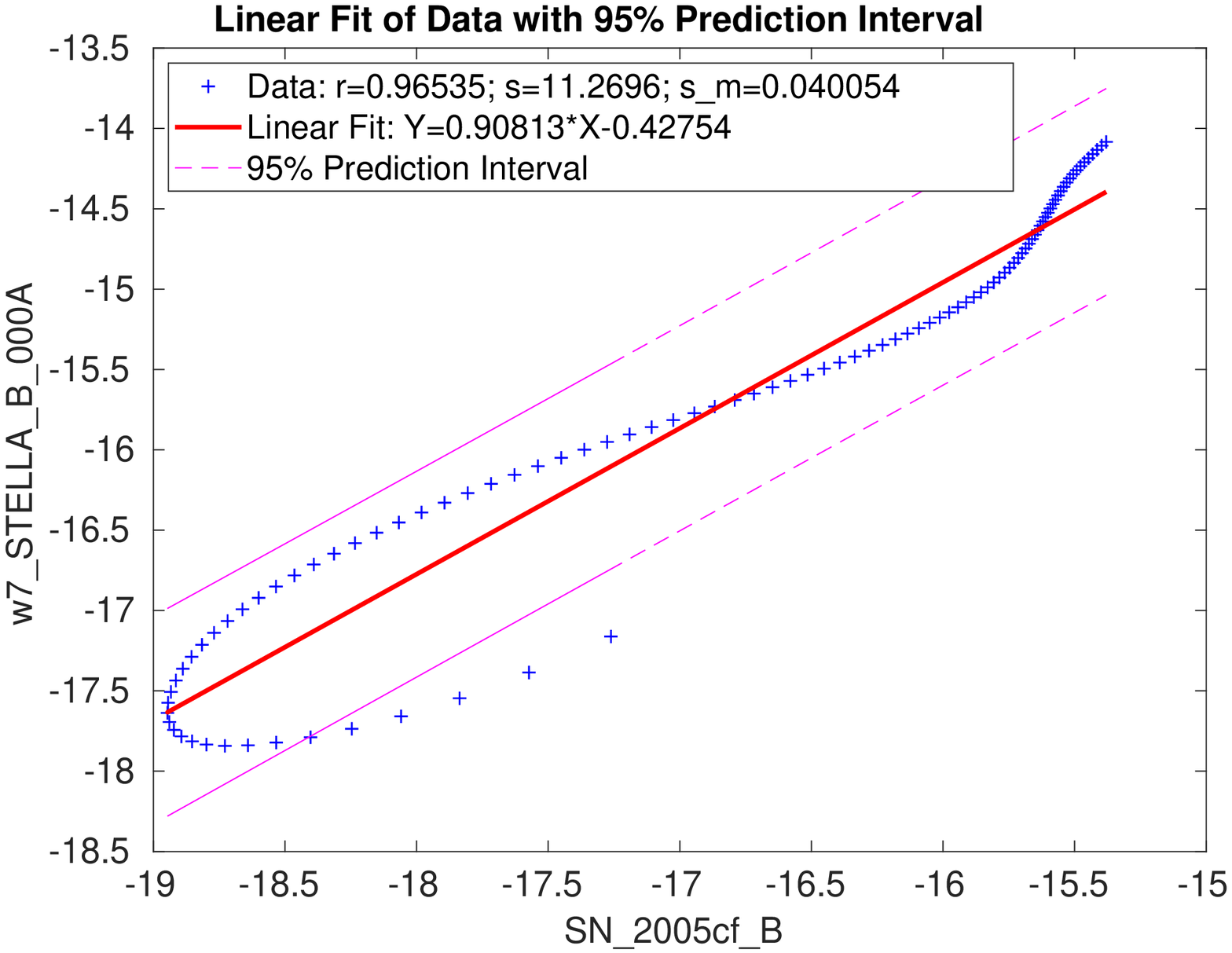}\hspace{5mm}
\includegraphics[width=0.3\textwidth]{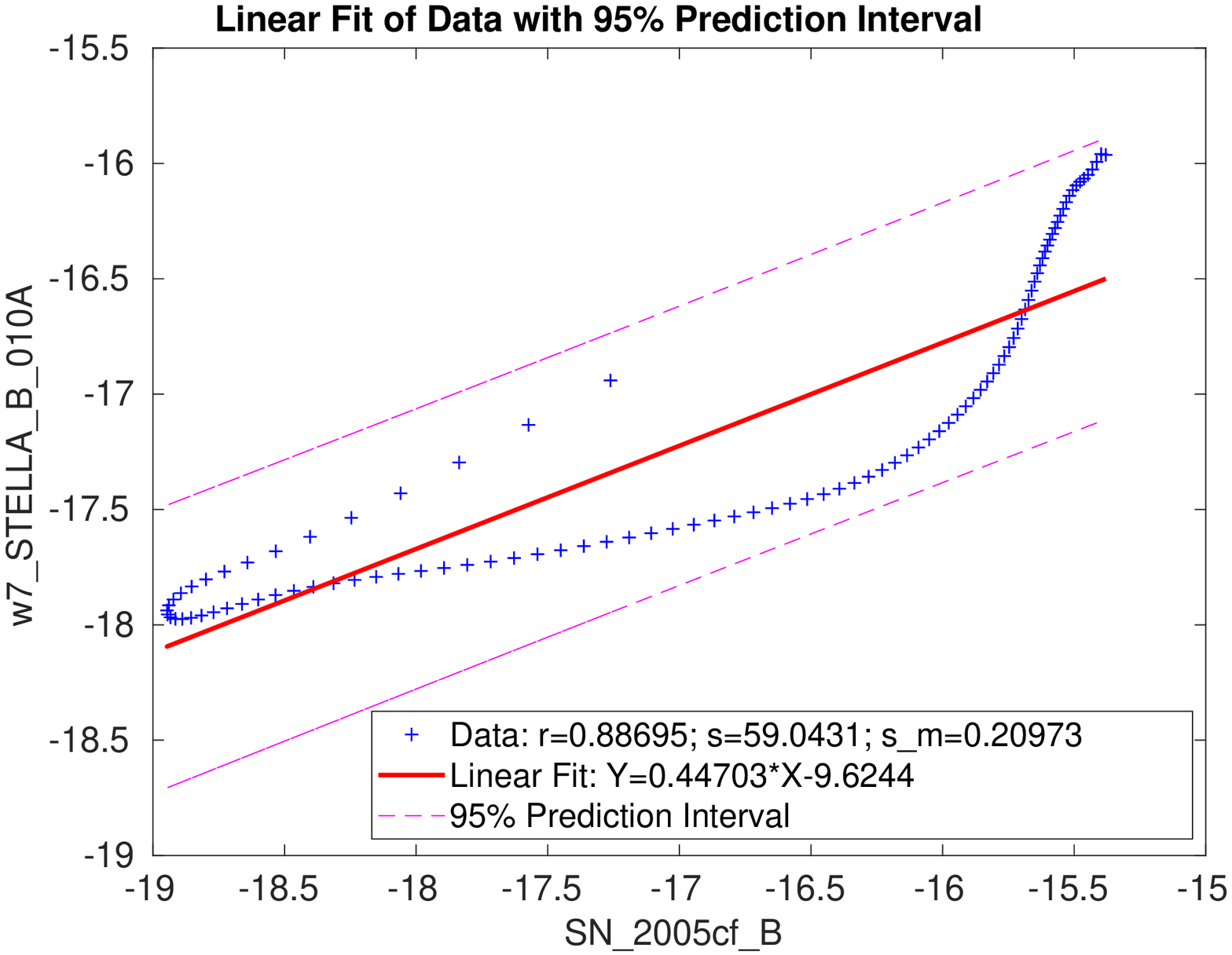}\hspace{5mm}
\includegraphics[width=0.3\textwidth]{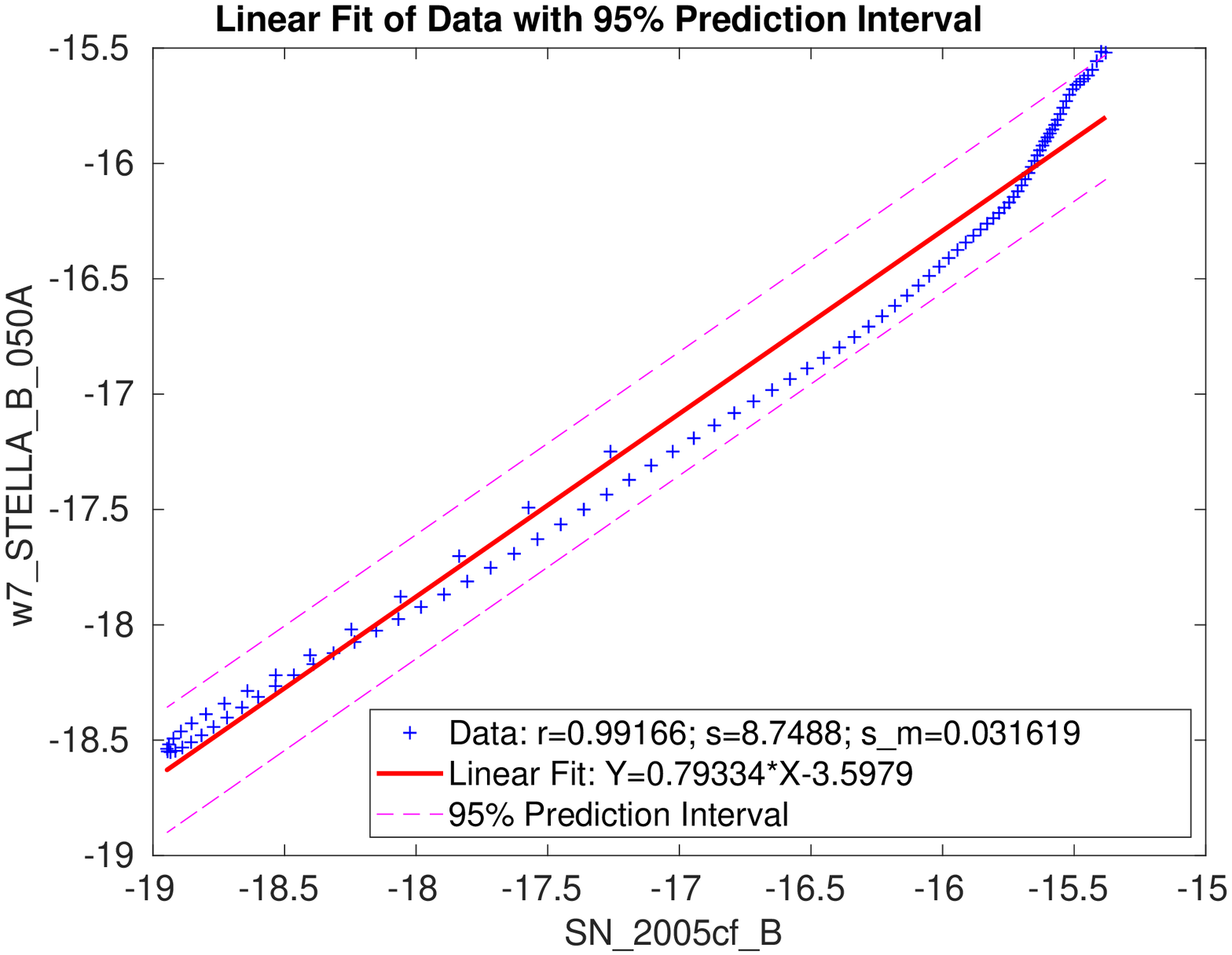}\\
\vspace{1mm}
\includegraphics[width=0.3\textwidth]{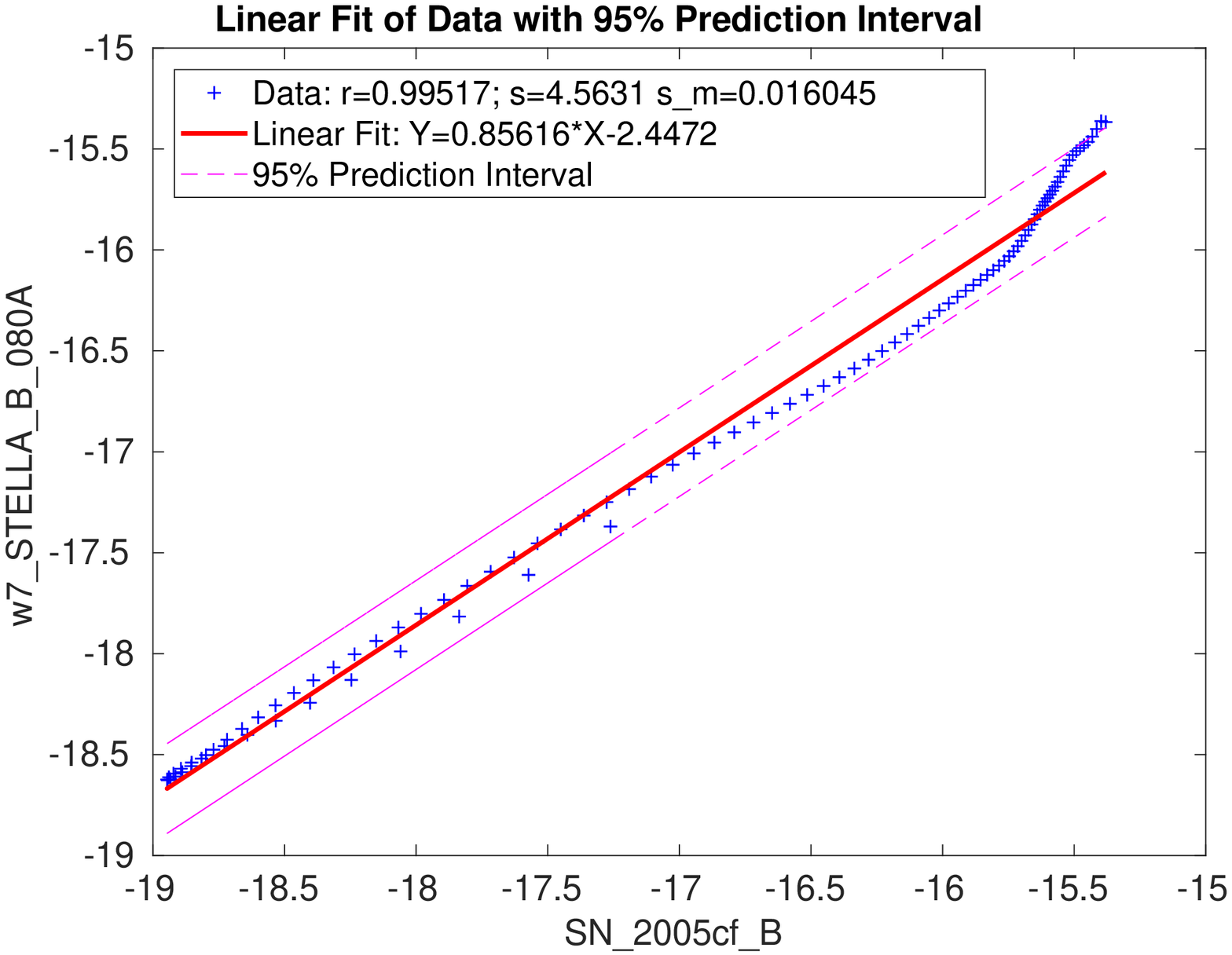}\hspace{5mm}
\includegraphics[width=0.3\textwidth]{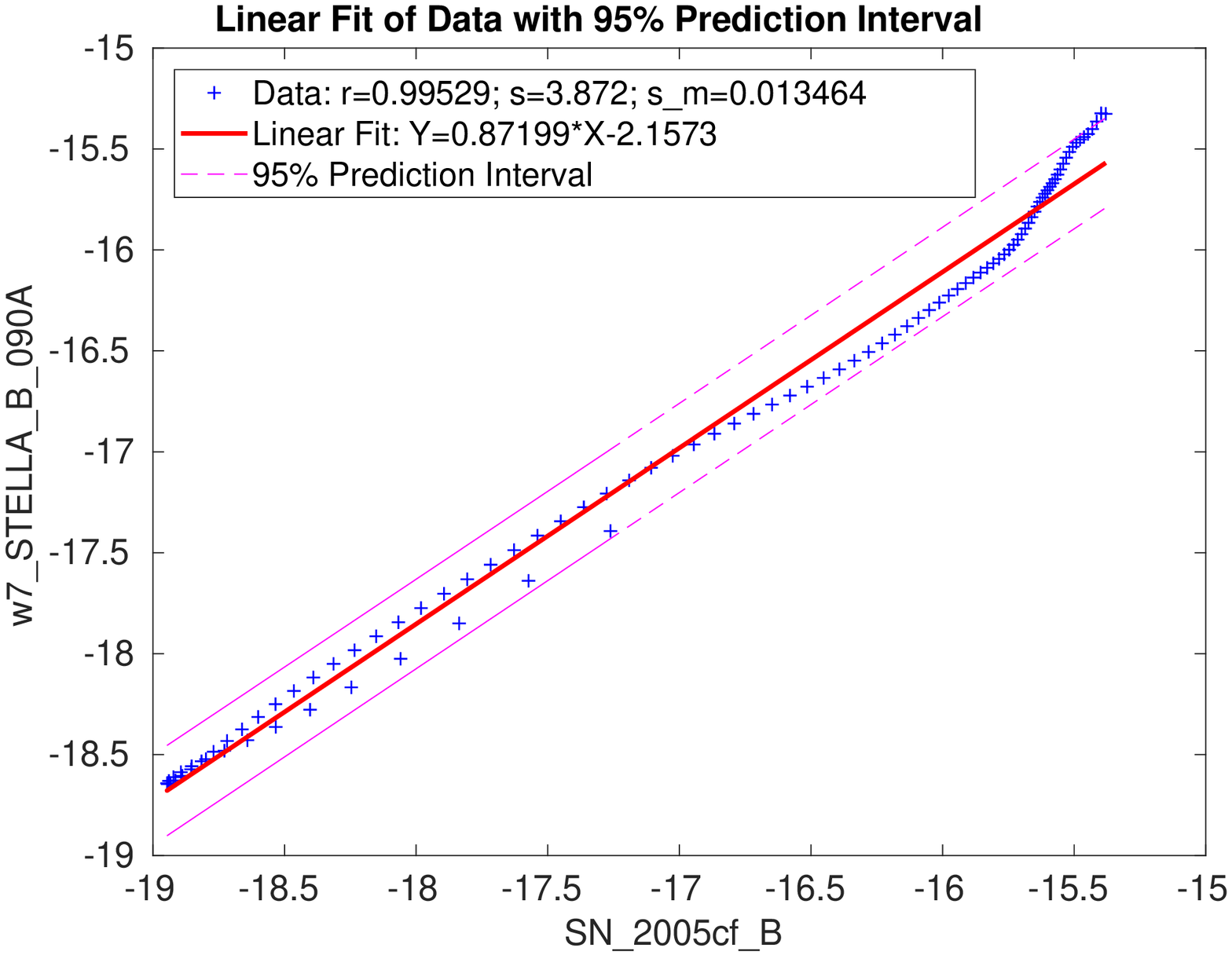}\hspace{5mm}
\includegraphics[width=0.3\textwidth]{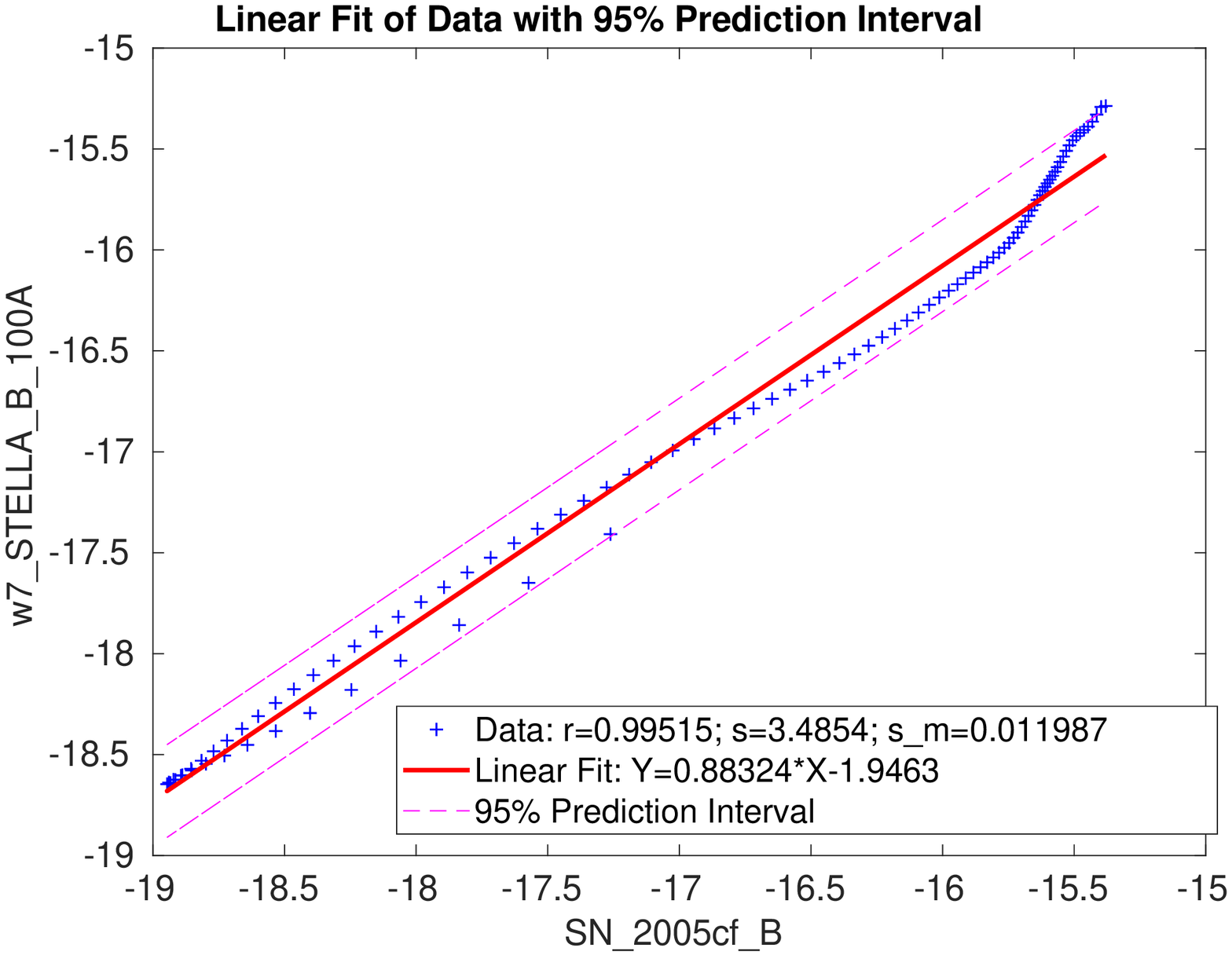}\\
\caption{The \emph{B}-band magnitudes of the W7 model
computed with STELLA using different assumed ratios between absorption and scattering, and observations of SN\,2005cf.}
\label{figure:2005cfcorr}
\end{figure*}

\begin{figure*}
\centering
\includegraphics[width=0.3\textwidth]{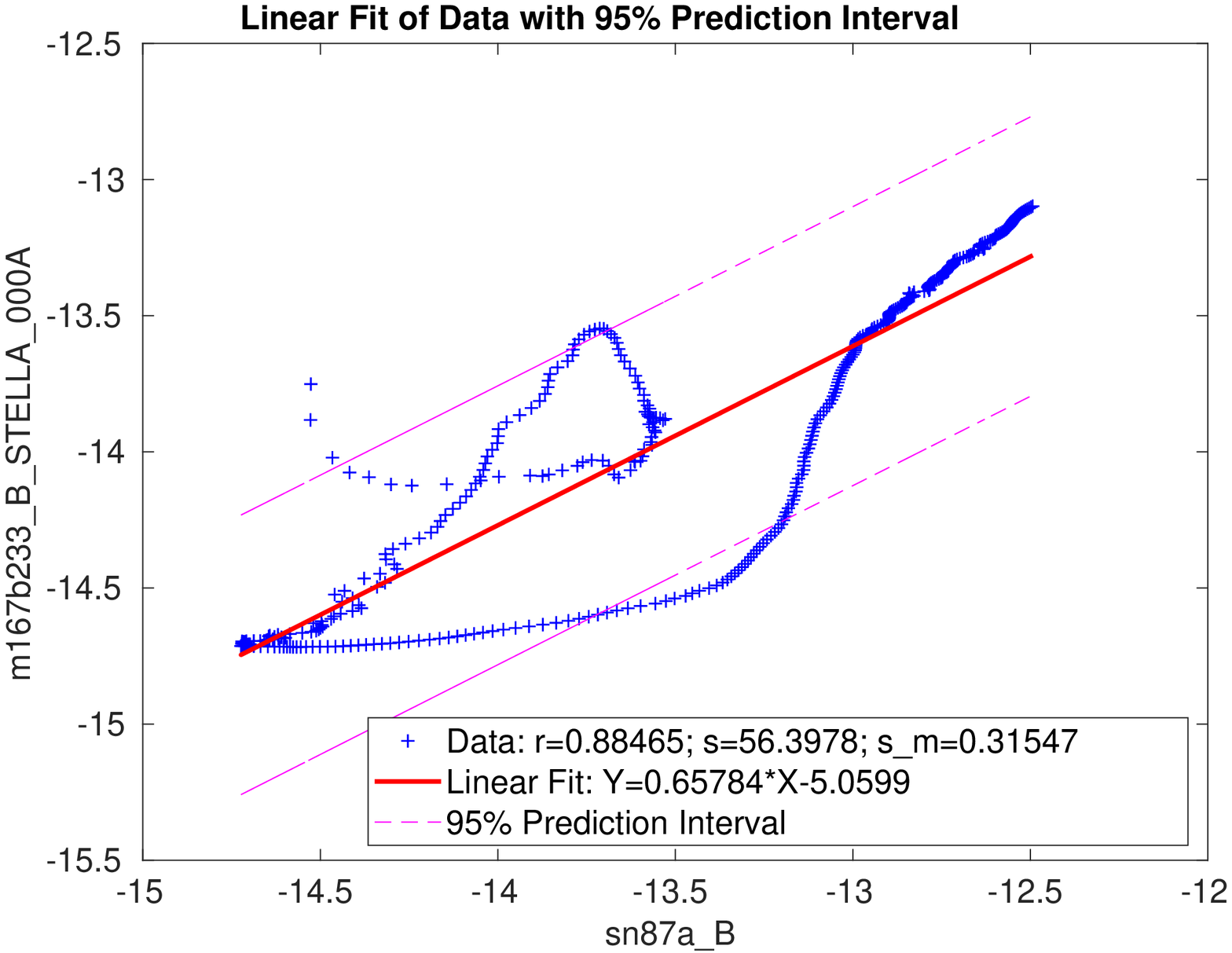}\hspace{5mm}
\includegraphics[width=0.3\textwidth]{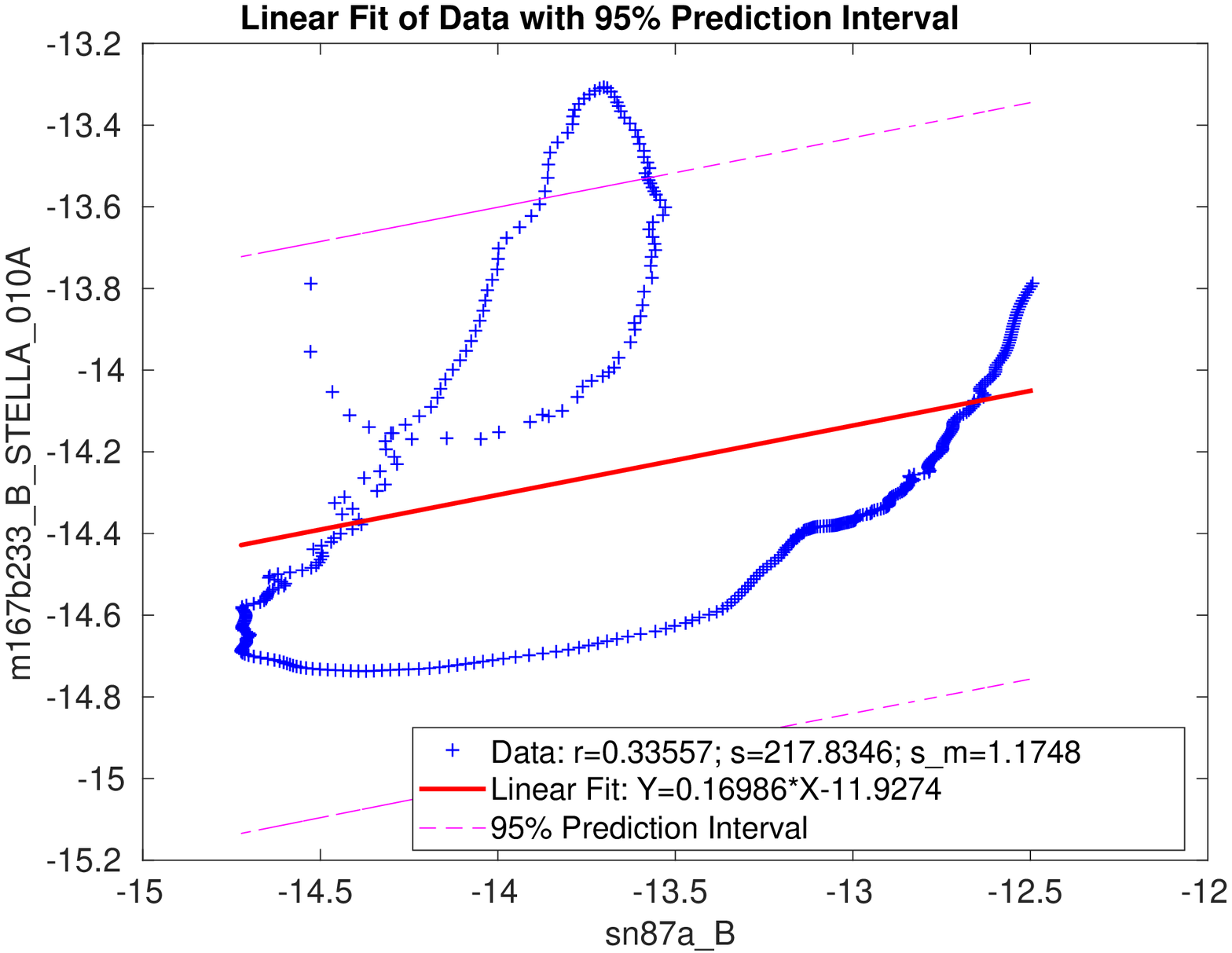}\hspace{5mm}
\includegraphics[width=0.3\textwidth]{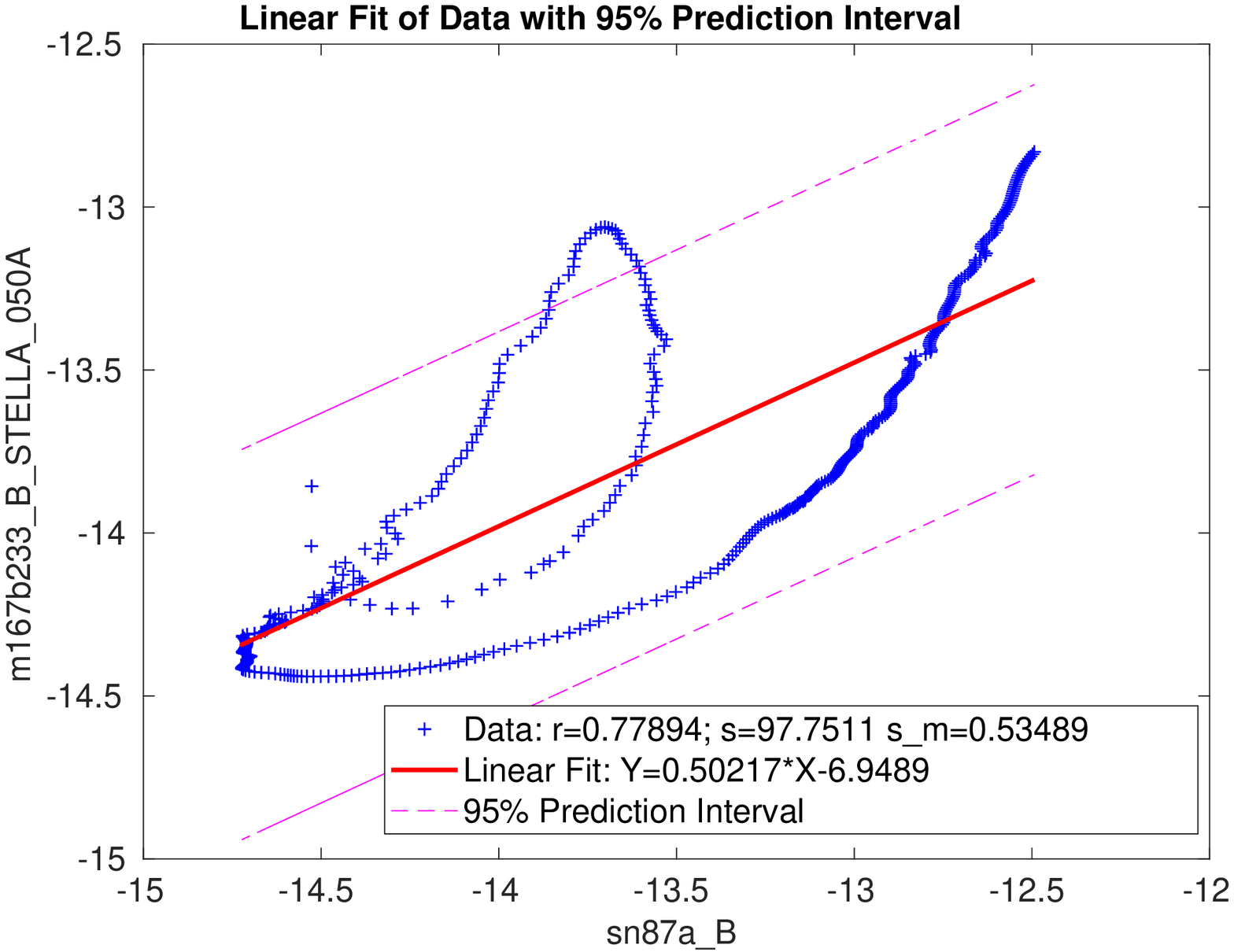}\\
\vspace{1mm}
\includegraphics[width=0.3\textwidth]{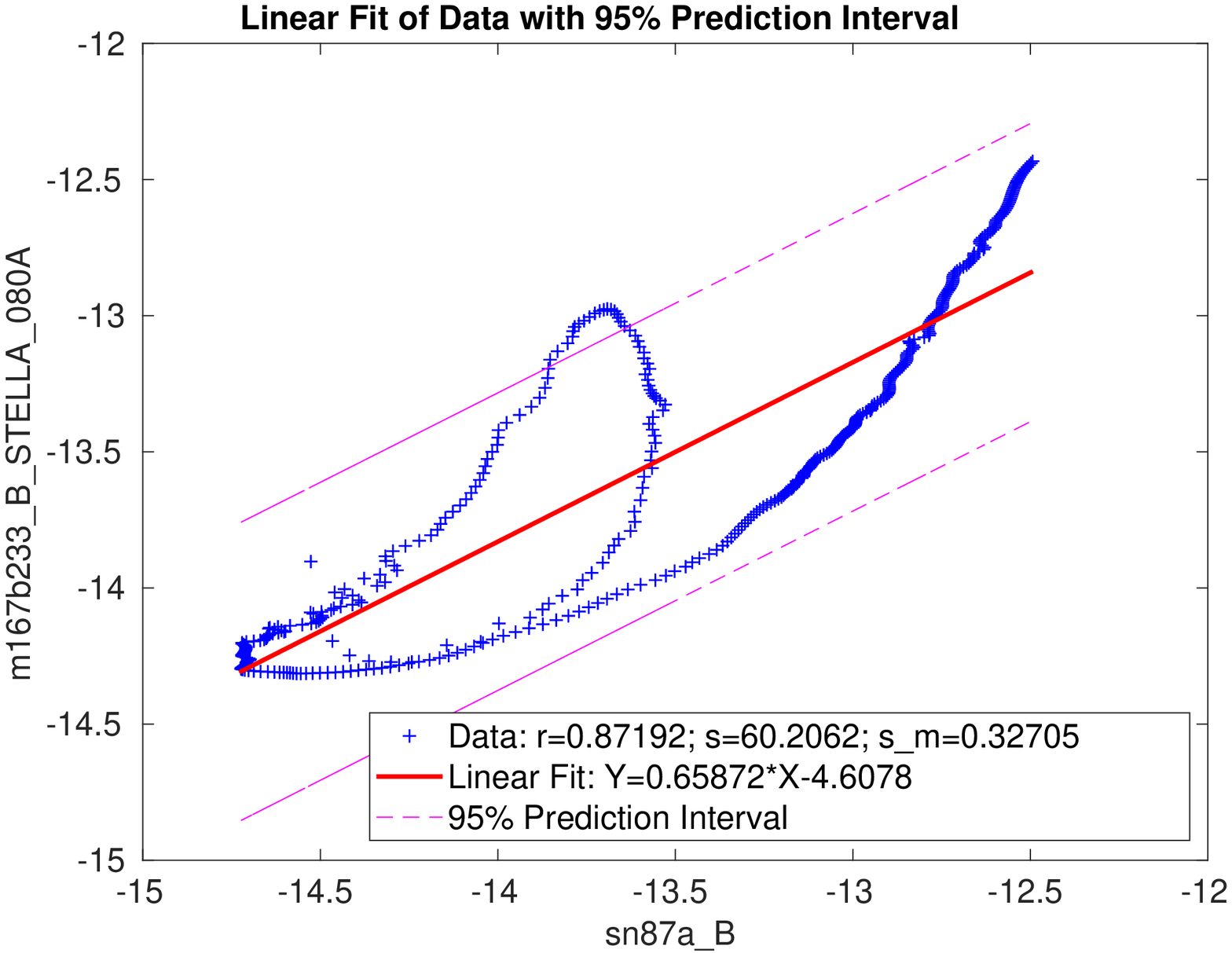}\hspace{5mm}
\includegraphics[width=0.3\textwidth]{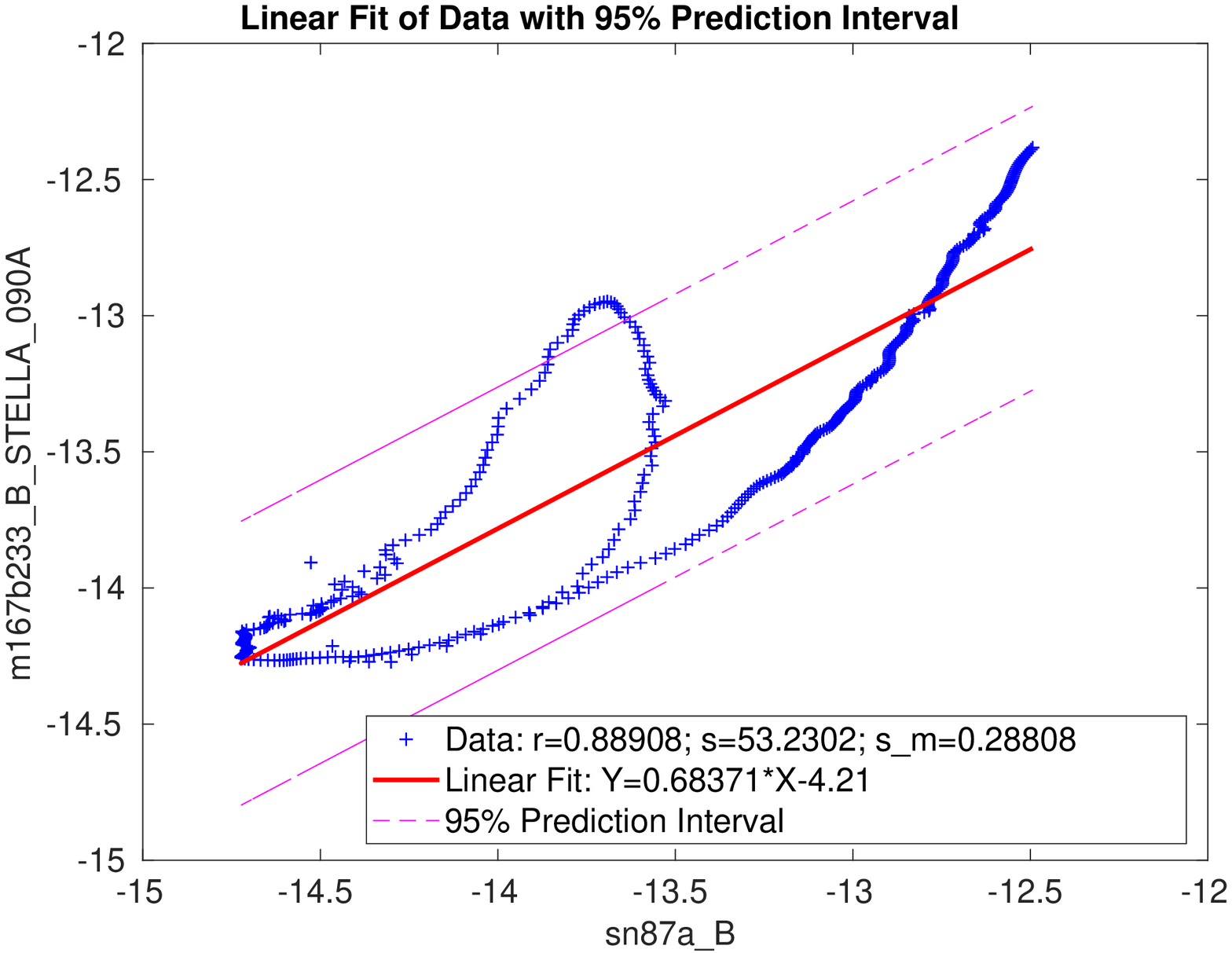}\hspace{5mm}
\includegraphics[width=0.3\textwidth]{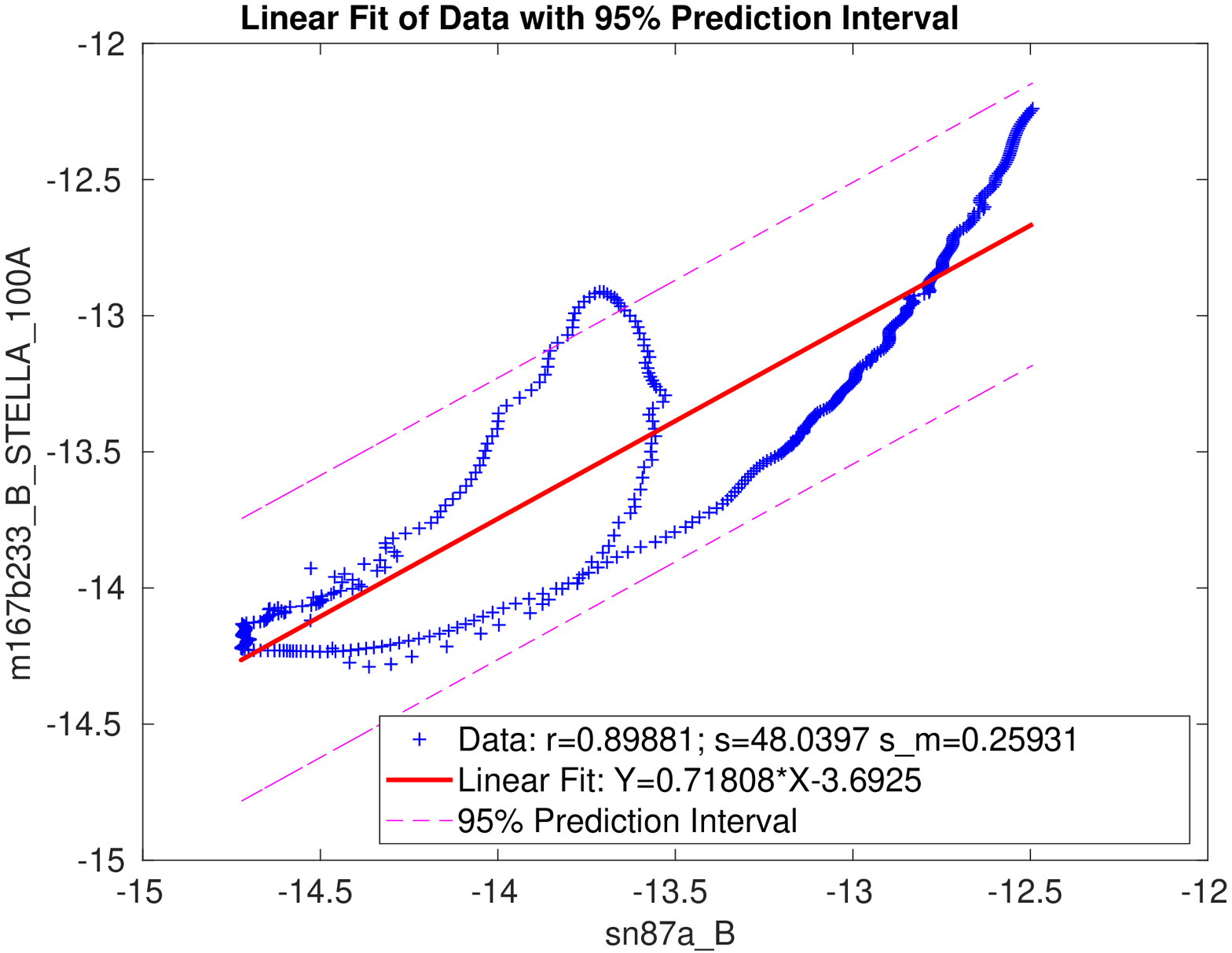}\\
\caption{The \emph{B}-band magnitudes for the 16-7b model computed
with STELLA using different ratios between absorption and scattering, and observations of SN\,1987A.}
\label{figure:87Acorr}
\end{figure*}

\begin{figure*}
\centering
\includegraphics[width=0.3\textwidth]{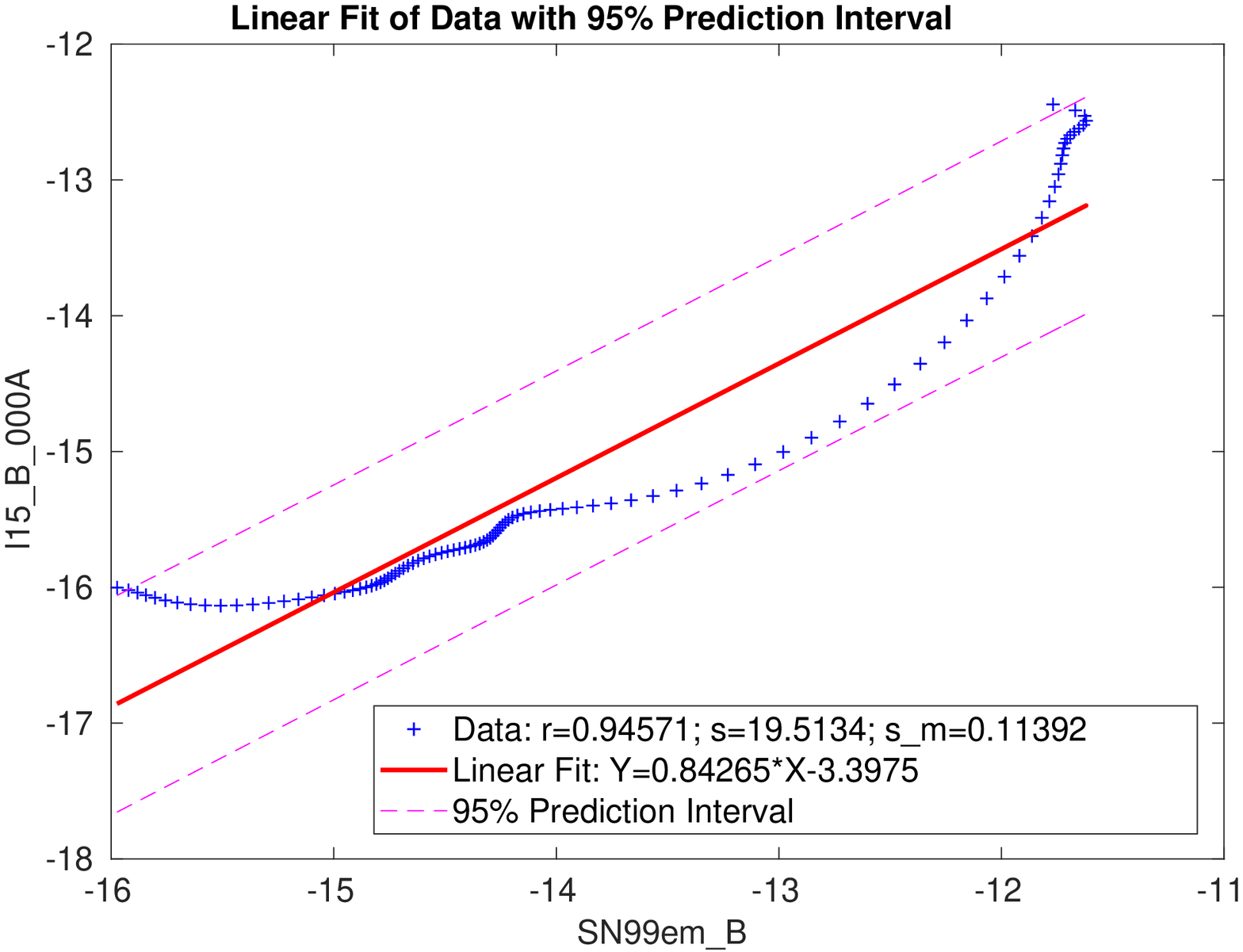}\hspace{5mm}
\includegraphics[width=0.3\textwidth]{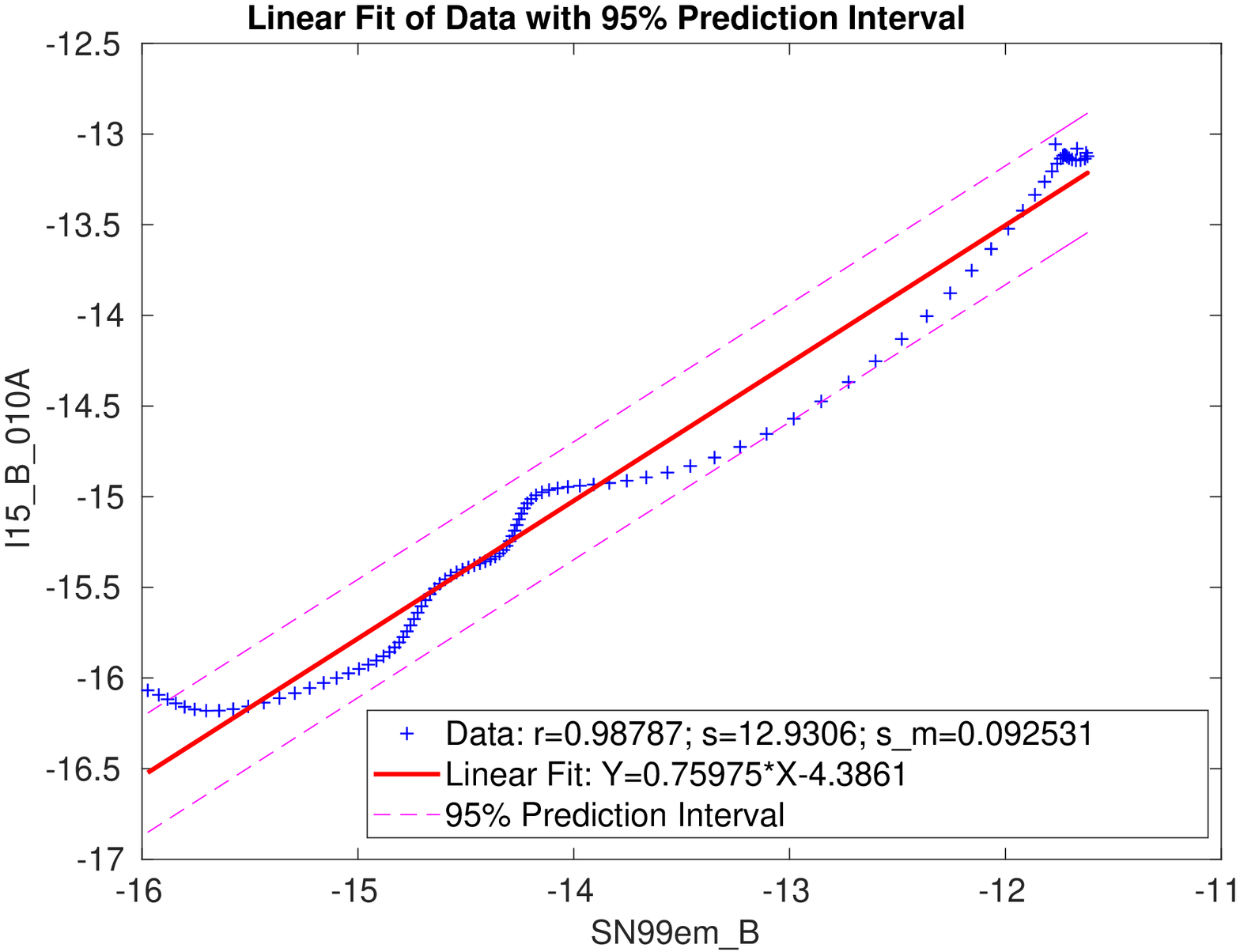}\hspace{5mm}
\includegraphics[width=0.3\textwidth]{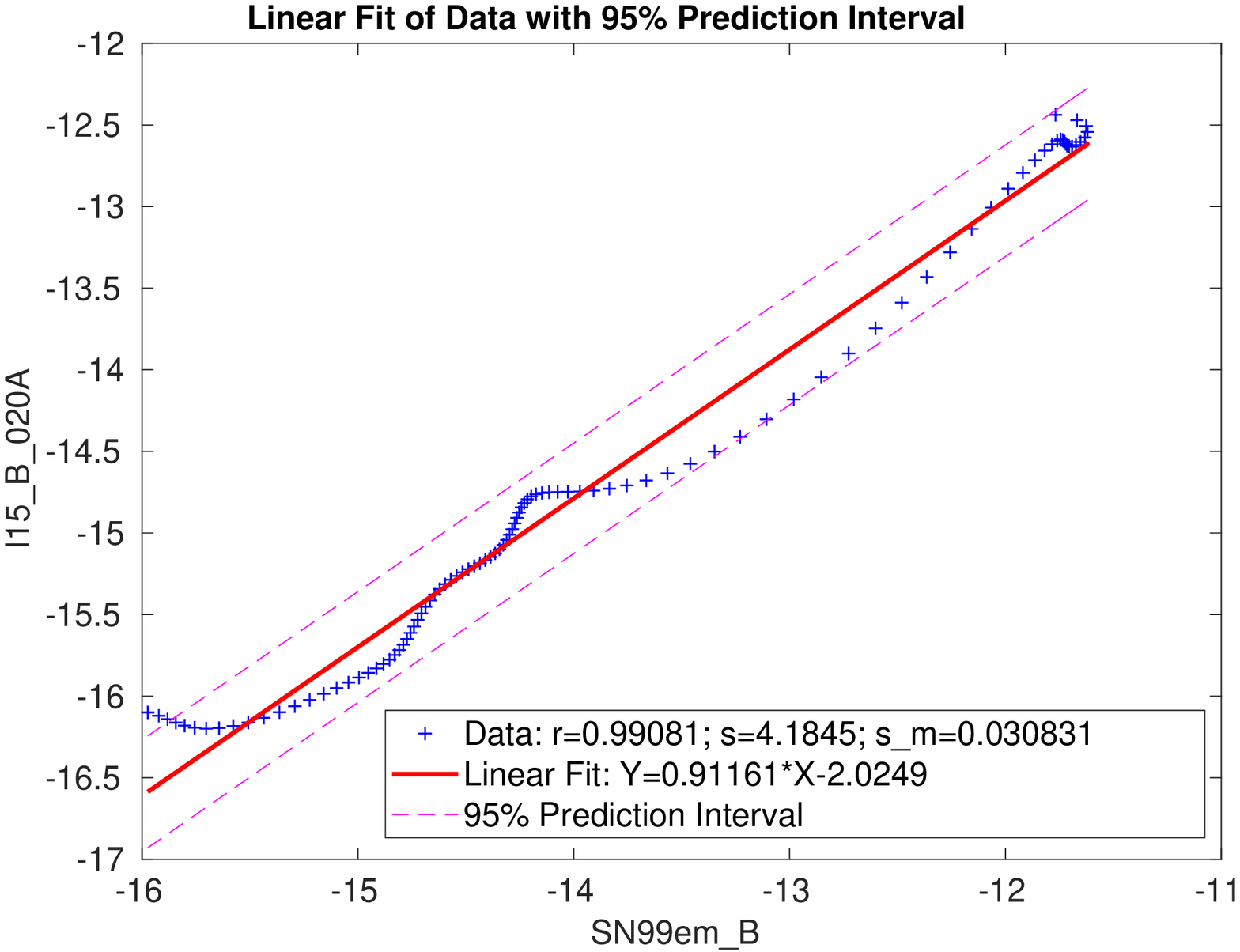}\\
\vspace{1mm}
\includegraphics[width=0.3\textwidth]{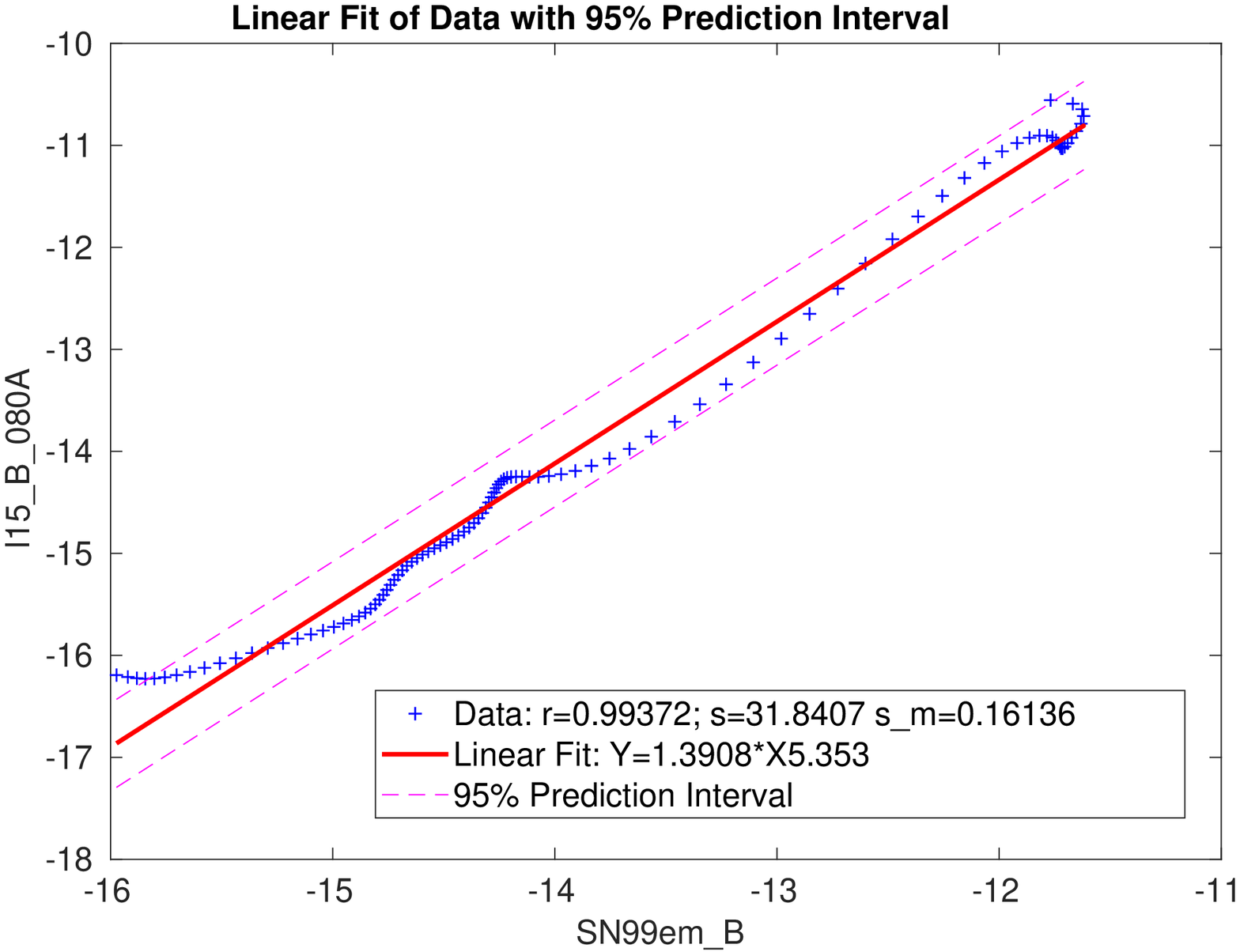}\hspace{5mm}
\includegraphics[width=0.3\textwidth]{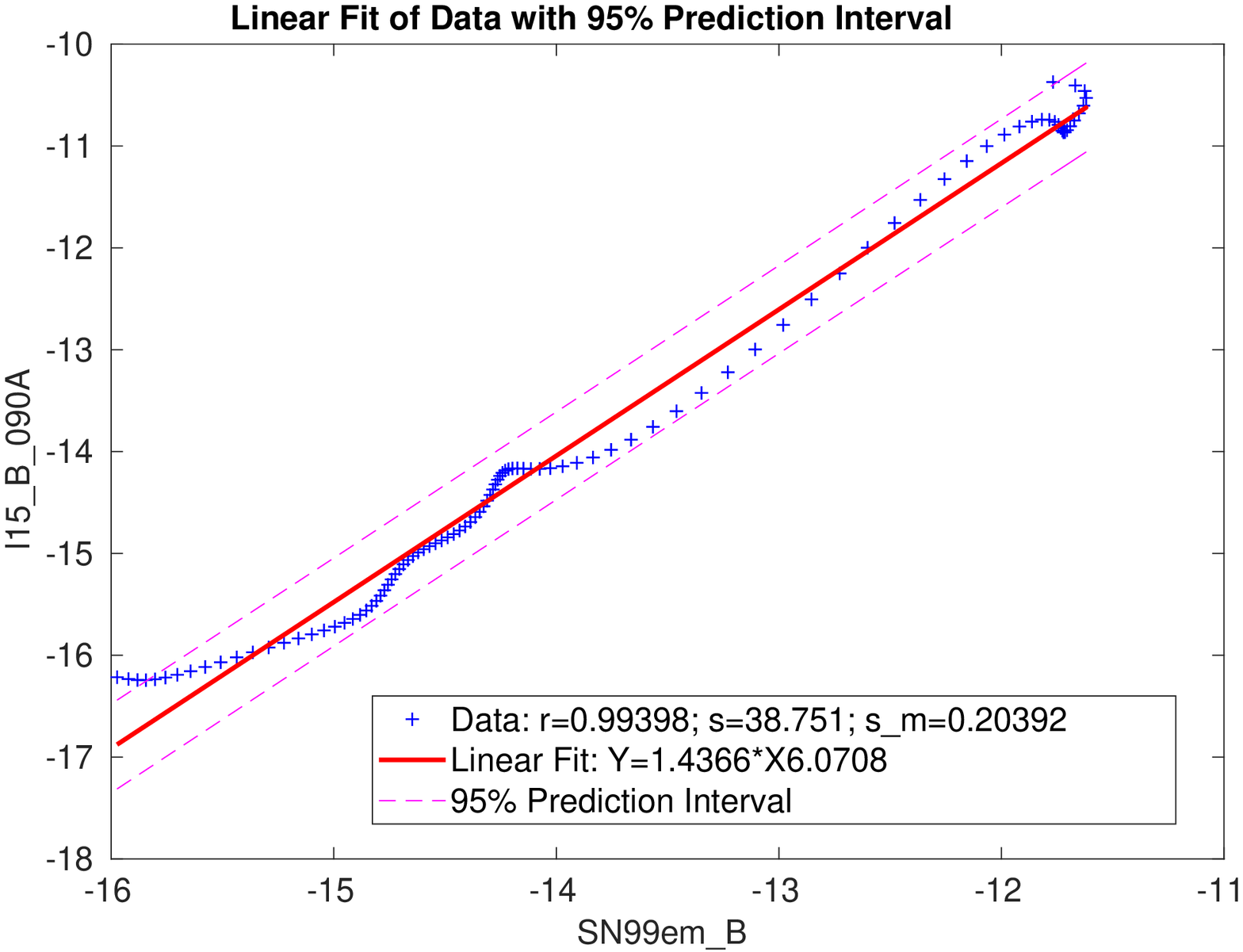}\hspace{5mm}
\includegraphics[width=0.3\textwidth]{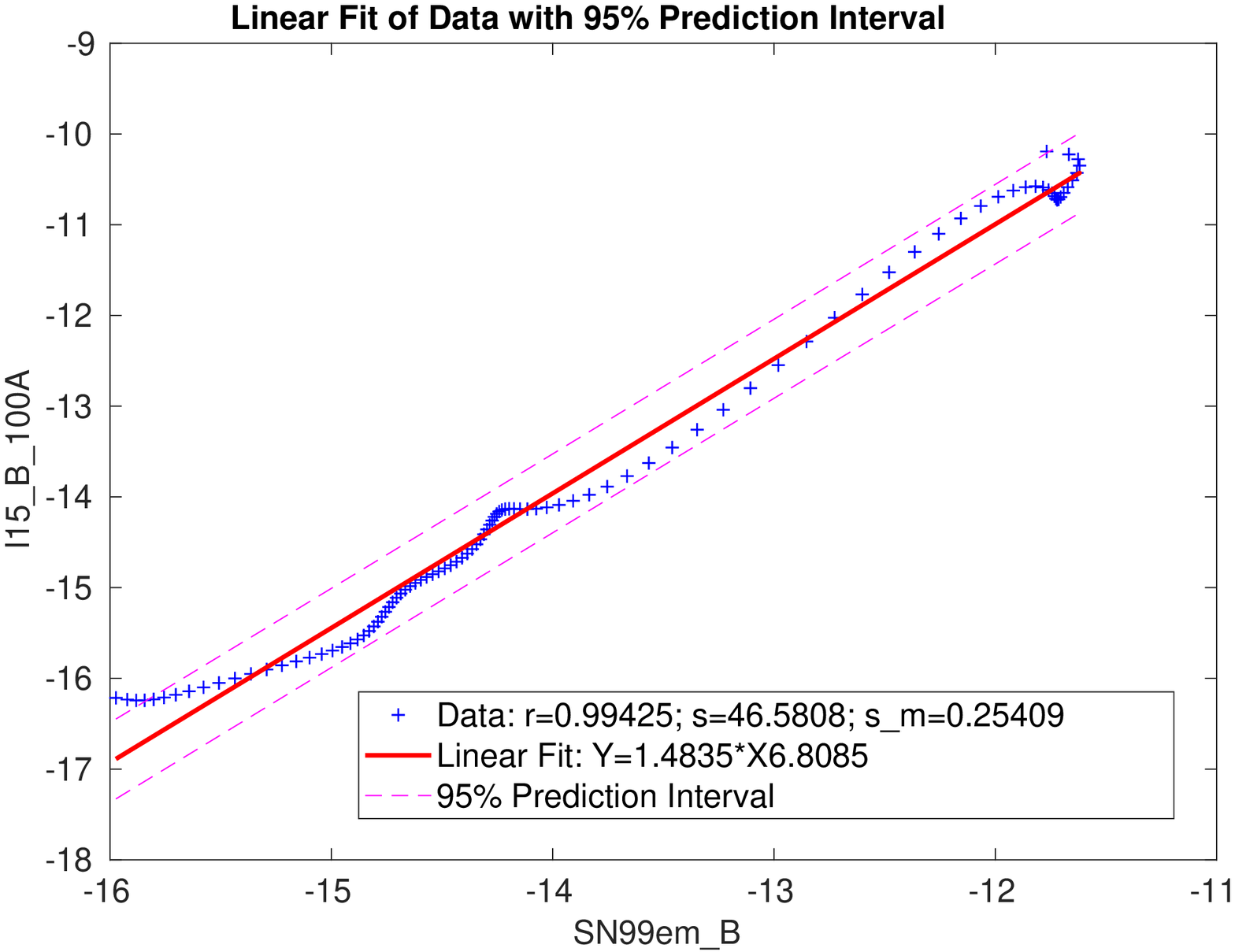}\\
\caption{The \emph{B}-band magnitudes for the L15 model
computed with STELLA using different ratios between absorption and scattering, and observations of SN\,1999em.}
\label{figure:1999emcorr}
\end{figure*}

In Figures~\ref{figure:2005cfcorr}, \ref{figure:87Acorr}, and \ref{figure:1999emcorr}, we
demonstrate the \emph{B}-band \verb|STELLA| magnitudes and corresponding magnitudes of the SN\,2005cf, SN\,1987A, and SN\,1999em.
Tables~\ref{table:coef3}, \ref{table:coef4}, and \ref{table:coef5} contain
calculated coefficients for the cases considered in the paper: the model W7
and SN\,2005cf, the model 16-7b and  SN\,1987A, and the model L15 and SN\,1999em.

\begin{table*}
\caption{The same as Table~\ref{table:coef1} for the W7 STELLA broad-band magnitudes and SN\,2005cf magnitudes.}
\label{table:coef3}
\begin{tabular}{l|l|l|l|l|l}
       &       & $A$ & $r$ & $s$  & $s_m$\\
\hline
U\underline{\,}sn2005cf & U\underline{\,}STELLA\underline{\,}000A & 0.81004 & 0.96627 & 17.38413 & 0.06189\\
U\underline{\,}sn2005cf & U\underline{\,}STELLA\underline{\,}010A & 0.67581 & 0.93254 & 36.48919 & 0.14668\\
U\underline{\,}sn2005cf & U\underline{\,}STELLA\underline{\,}050A & 1.05999 & 0.99128 & 4.92997 & 0.01490\\
U\underline{\,}sn2005cf & U\underline{\,}STELLA\underline{\,}080A & 1.13471 & 0.99130 & 8.58731 & 0.02674\\
U\underline{\,}sn2005cf & U\underline{\,}STELLA\underline{\,}090A & 1.15321 & 0.99087 & 10.11125&0.03232\\
U\underline{\,}sn2005cf & U\underline{\,}STELLA\underline{\,}100A & 1.16688 & 0.99045 & 11.40321 & 0.03729\\
\hline
B\underline{\,}sn2005cf & B\underline{\,}STELLA\underline{\,}000A & 0.90813 & 0.96535 & 11.26962 & 0.04005\\
B\underline{\,}sn2005cf & B\underline{\,}STELLA\underline{\,}010A & 0.44703 & 0.88695 & 59.04310 & 0.20973\\
B\underline{\,}sn2005cf & B\underline{\,}STELLA\underline{\,}050A & 0.79334 & 0.99166 & 8.74879 & 0.03162\\
B\underline{\,}sn2005cf & B\underline{\,}STELLA\underline{\,}080A & 0.85616& 0.99517& 4.56309& 0.01604\\
B\underline{\,}sn2005cf & B\underline{\,}STELLA\underline{\,}090A & 0.87199 & 0.99529 & 3.87200 & 0.01346\\
B\underline{\,}sn2005cf & B\underline{\,}STELLA\underline{\,}100A & 0.88324 & 0.99515 & 3.48541 & 0.01199\\
\hline
V\underline{\,}sn2005cf & V\underline{\,}STELLA\underline{\,}000A &  1.22312 & 0.93853 & 20.54001 & 0.09563\\
V\underline{\,}sn2005cf & V\underline{\,}STELLA\underline{\,}010A & 0.30969 & 0.82297 & 42.53423& 0.13506\\
V\underline{\,}sn2005cf & V\underline{\,}STELLA\underline{\,}050A & 0.65986 & 0.99167 & 10.02148 & 0.03177\\
V\underline{\,}sn2005cf & V\underline{\,}STELLA\underline{\,}080A & 0.72819& 0.99885 & 6.11644 & 0.01934\\
V\underline{\,}sn2005cf & V\underline{\,}STELLA\underline{\,}090A & 0.74629 & 0.99915 & 5.31970 & 0.01679\\
V\underline{\,}sn2005cf & V\underline{\,}STELLA\underline{\,}100A &0.76039 & 0.99922 &  4.75023 & 0.01496 \\
\hline
R\underline{\,}sn2005cf & R\underline{\,}STELLA\underline{\,}000A & 1.24223 & 0.91316 & 25.27996 & 0.10852\\
R\underline{\,}sn2005cf & R\underline{\,}STELLA\underline{\,}010A & 0.24682 & 0.55332 & 48.70329 & 0.15306\\
R\underline{\,}sn2005cf & R\underline{\,}STELLA\underline{\,}050A & 0.60890 & 0.88714 & 17.48942 & 0.05509\\
R\underline{\,}sn2005cf & R\underline{\,}STELLA\underline{\,}080A & 0.67632 & 0.92815 & 12.31342& 0.03897 \\
R\underline{\,}sn2005cf & R\underline{\,}STELLA\underline{\,}090A & 0.69377 & 0.93631 & 11.15055 & 0.03535\\
R\underline{\,}sn2005cf & R\underline{\,}STELLA\underline{\,}100A & 0.70649 & 0.94276 & 10.26031 & 0.03257\\
\hline
I\underline{\,}sn2005cf & I\underline{\,}STELLA\underline{\,}000A & 1.45465 & 0.90383 & 32.08675 & 0.12960\\
I\underline{\,}sn2005cf & I\underline{\,}STELLA\underline{\,}010A & 0.09844 & 0.28903 & 43.30838 & 0.13566\\
I\underline{\,}sn2005cf & I\underline{\,}STELLA\underline{\,}050A & 0.54445 & 0.82089 & 16.53865 & 0.05180\\
I\underline{\,}sn2005cf & I\underline{\,}STELLA\underline{\,}080A & 0.61584 & 0.86225 & 13.11988 & 0.04129\\
I\underline{\,}sn2005cf & I\underline{\,}STELLA\underline{\,}090A & 0.63418 & 0.87236 & 12.25684 & 0.03864\\
I\underline{\,}sn2005cf & I\underline{\,}STELLA\underline{\,}100A & 0.64790 & 0.87824 & 11.70627 & 0.03694\\
\end{tabular}
\end{table*}


\begin{table*}
\caption{The same as Table~\ref{table:coef1} for the 16-7b STELLA broad-band magnitudes and SN\,1987A magnitudes.}
\label{table:coef4}
\begin{tabular}{l|l|l|l|l|l}
       &       & $A$ & $r$ & $s$  & $s_m$\\
\hline
U\underline{\,}87A&U\underline{\,}STELLA\underline{\,}000A &0.06837&0.13050&317.80879&2.28892\\
U\underline{\,}87A&U\underline{\,}STELLA\underline{\,}010A &0.08695&0.13709&343.13217&2.44217\\
U\underline{\,}87A&U\underline{\,}STELLA\underline{\,}050A &0.76728&0.85858&73.76258&0.56205\\
U\underline{\,}87A&U\underline{\,}STELLA\underline{\,}080A &1.03185&0.91171&60.67765&0.43050\\
U\underline{\,}87A&U\underline{\,}STELLA\underline{\,}090A &1.08594&0.92123&60.80450&0.43279\\
U\underline{\,}87A&U\underline{\,}STELLA\underline{\,}100A &1.13518&0.92595&64.97415&0.46718\\
\hline
B\underline{\,}87A&B\underline{\,}STELLA\underline{\,}000A &0.65784 &0.88465 &56.3978 &0.31457\\
B\underline{\,}87A&B\underline{\,}STELLA\underline{\,}010A &0.16986 &0.33557 &217.8346 &1.1748\\
B\underline{\,}87A&B\underline{\,}STELLA\underline{\,}050A &0.50217 &0.77894 &97.7511 &0.53489\\
B\underline{\,}87A&B\underline{\,}STELLA\underline{\,}080A &0.65872 &0.87192 &60.2062 &0.32705\\
B\underline{\,}87A&B\underline{\,}STELLA\underline{\,}090A &0.68371 &0.88908 &53.2302 &0.28808\\
B\underline{\,}87A&B\underline{\,}STELLA\underline{\,}100A &0.71808 &0.89881 &48.0397 &0.25931\\
\hline
V\underline{\,}87A&V\underline{\,}STELLA\underline{\,}000A &0.79803 &0.92585 &35.4874 &0.15782\\
V\underline{\,}87A&V\underline{\,}STELLA\underline{\,}010A &0.45557 &0.74768 &111.1707 & 0.50002\\
V\underline{\,}87A&V\underline{\,}STELLA\underline{\,}050A &0.58864 &0.85189 &72.5256 &0.32627\\
V\underline{\,}87A&V\underline{\,}STELLA\underline{\,}080A &0.67421 &0.90059 &51.2285 &0.22959\\
V\underline{\,}87A&V\underline{\,}STELLA\underline{\,}090A &0.68007 &0.91032 &47.8315 &0.21402\\
V\underline{\,}87A&V\underline{\,}STELLA\underline{\,}100A &0.69792 &0.91697 &44.3254 &0.19796\\
\hline
R\underline{\,}87A&R\underline{\,}STELLA\underline{\,}000A&0.91404&0.92960&23.77577&0.09832\\
R\underline{\,}87A&R\underline{\,}STELLA\underline{\,}010A&0.82889&0.92540&24.76832&0.10066\\
R\underline{\,}87A&R\underline{\,}STELLA\underline{\,}050A&0.81238&0.92252&25.83081&0.10519\\
R\underline{\,}87A&R\underline{\,}STELLA\underline{\,}080A&0.84719&0.93368&22.09643&0.08996\\
R\underline{\,}87A&R\underline{\,}STELLA\underline{\,}090A&0.84185&0.93614&21.42282&0.08721\\
R\underline{\,}87A&R\underline{\,}STELLA\underline{\,}100A&0.84607&0.93794&20.83242&0.08481\\
\hline
I\underline{\,}87A&I\underline{\,}STELLA\underline{\,}000A&0.99412&0.94056&20.47483&0.08229\\
I\underline{\,}87A&I\underline{\,}STELLA\underline{\,}010A&0.96658&0.96296&11.81486&0.04632\\
I\underline{\,}87A&I\underline{\,}STELLA\underline{\,}050A&0.88718&0.94229&17.80415&0.06965\\
I\underline{\,}87A&I\underline{\,}STELLA\underline{\,}080A&0.90315&0.94723&16.32774&0.06381\\
I\underline{\,}87A&I\underline{\,}STELLA\underline{\,}090A&0.90073&0.94690&16.42541&0.06422\\
I\underline{\,}87A&I\underline{\,}STELLA\underline{\,}100A&0.89997&0.94716&16.34396&0.06393\\
\end{tabular}
\end{table*}


\begin{table*}
\caption{The same as Table~\ref{table:coef1} for the L15 STELLA broad-band magnitudes and SN\,1999em magnitudes.}
\label{table:coef5}
\begin{tabular}{l|l|l|l|l|l}
       &       & $A$ & $r$ & $s$  & $s_m$\\
\hline
U\underline{\,}sn99em& U\underline{\,}STELLA\underline{\,}L15\underline{\,}000A&0.30067&0.97527&80.26059&0.44725\\
U\underline{\,}sn99em& U\underline{\,}STELLA\underline{\,}L15\underline{\,}010A&0.67449&0.98157&20.02893&0.13150\\
U\underline{\,}sn99em& U\underline{\,}STELLA\underline{\,}L15\underline{\,}020A&0.76093&0.97187&14.82002&0.09496\\
U\underline{\,}sn99em& U\underline{\,}STELLA\underline{\,}L15\underline{\,}080A&0.93246&0.98199&5.97510&0.03312\\
U\underline{\,}sn99em& U\underline{\,}STELLA\underline{\,}L15\underline{\,}090A&0.95300&0.98158&5.95111&0.03151\\
U\underline{\,}sn99em& U\underline{\,}STELLA\underline{\,}L15\underline{\,}100A&0.97224&0.98208&5.78599&0.02953\\
\hline
B\underline{\,}sn99em&B\underline{\,}STELLA\underline{\,}L15\underline{\,}000A&0.84265&0.94571&19.51341&0.11392\\
B\underline{\,}sn99em&B\underline{\,}STELLA\underline{\,}L15\underline{\,}010A&0.75975&0.98787&12.93062&0.09253\\
B\underline{\,}sn99em&B\underline{\,}STELLA\underline{\,}L15\underline{\,}020A&0.91161&0.99081&4.18450&0.03083\\
B\underline{\,}sn99em&B\underline{\,}STELLA\underline{\,}L15\underline{\,}080A&1.39078&0.99372&31.84070&0.16136\\
B\underline{\,}sn99em&B\underline{\,}STELLA\underline{\,}L15\underline{\,}090A&1.43664&0.99398&38.75098&0.20392\\
B\underline{\,}sn99em&B\underline{\,}STELLA\underline{\,}L15\underline{\,}100A&1.48351&0.99425&46.58084&0.25409\\
\hline
V\underline{\,}sn99em&V\underline{\,}STELLA\underline{\,}L15\underline{\,}000A&1.13095&0.98045&7.21903&0.02631\\
V\underline{\,}sn99em&V\underline{\,}STELLA\underline{\,}L15\underline{\,}010A&0.86758&0.98447&4.36066&0.02386\\
V\underline{\,}sn99em&V\underline{\,}STELLA\underline{\,}L15\underline{\,}020A&0.95116&0.98997&2.18841&0.01139\\
V\underline{\,}sn99em&V\underline{\,}STELLA\underline{\,}L15\underline{\,}080A&1.26760&0.99398&9.58007&0.03485\\
V\underline{\,}sn99em&V\underline{\,}STELLA\underline{\,}L15\underline{\,}090A&1.30034&0.99346&11.83005&0.04448\\
V\underline{\,}sn99em&V\underline{\,}STELLA\underline{\,}L15\underline{\,}100A&1.33826&0.99367&14.437937&0.05595\\
\end{tabular}
\end{table*}








%


\bsp	
\label{lastpage}
\end{document}